\documentclass[11pt]{article}
\usepackage{amsmath,amssymb}
\usepackage{graphicx,side}
\usepackage{psfrag}
\normalsize

\makeatletter

 \@addtoreset{equation}{section}
\makeatother

\newcommand{\Bracket}[2]{\langle \, {#1} \, {#2}\, \rangle}
\newcommand{\Slash}[1]{\ooalign{\hfil/\hfil\crcr$#1$}}

\newcommand{\lpd}[1]{\frac{\partial}{\partial{#1}}}
\newcommand{\del}{\partial}
\newcommand{\half}{\frac{1}{2}}
\newcommand{\mbf}{\boldsymbol}

\newcommand{\wt}{\widetilde}

\newcommand{\ol}{\overline}

\newcommand{\dps}{\displaystyle}
\newcommand{\vs}[1]{\vspace{#1 mm}}
\newcommand{\thb}{\ol{\theta}}
\newcommand{\hs}[1]{\hspace{#1 mm}}

\begin{document}

\allowdisplaybreaks{


\setcounter{page}{0}

\begin{titlepage}

\begin{flushright}
OU-HET 366\\
TIT/HEP-457\\
hep-th/0010272\\
October 2000
\end{flushright}
\bigskip

\begin{center}
{\LARGE\bf
Large-$\mbf{N}$ Limit of 
$\mbf{{\cal N}=2}$ Supersymmetric $\mbf{Q^{N}}$ Model\\ 

in Two Dimensions
}
\vs{10}

\bigskip
{\renewcommand{\thefootnote}{\fnsymbol{footnote}}
{\large\bf Kiyoshi Higashijima$^a$\footnote{
     E-mail: higashij@phys.sci.osaka-u.ac.jp},
 Tetsuji Kimura$^a$\footnote{
     E-mail: t-kimura@het.phys.sci.osaka-u.ac.jp},
 Muneto Nitta$^b$\footnote{
     E-mail: nitta@th.phys.titech.ac.jp}
 and Makoto Tsuzuki$^a$\footnote{
     E-mail: tsuzki@het.phys.sci.osaka-u.ac.jp}
}}

\setcounter{footnote}{0}
\bigskip

{\small \it
$^a$Department of Physics,
Graduate School of Science, Osaka University,\\
Toyonaka, Osaka 560-0043, Japan\\
$^b$Department of Physics, Tokyo Institute of Technology,\\ 
Oh-okayama, Meguro, Tokyo 152-8551, Japan\\
}
\end{center}
\bigskip

\begin{abstract}
We investigate non-perturbative structures of 
the two-dimensional ${\cal N}=2$ supersymmetric nonlinear sigma model 
on the quadric surface 
$Q^{N-2}({\bf C}) = SO(N)/SO(N-2)\times U(1)$, 
which is a Hermitian symmetric space, 
and therefore K\"{a}hler,   
by using the auxiliary field and large-$N$ methods.   
This model contains two kinds of 
non-perturbatively stable vacua;  
one of them is the same vacuum as that of 
supersymmetric ${\bf C}P^{N-1}$ model, 
and the other is a new kind of vacuum,  
which has not yet been known to exist  
in two-dimensional nonlinear sigma models, the Higgs phase.  
We show that both of these vacua are asymptotically free. 
Although symmetries are broken in these vacua,   
there appear no massless Nambu-Goldstone bosons,  
in agreement with Coleman's theorem, 
due to the existence of two different mechanisms in these vacua, 
the Schwinger and the Higgs mechanisms.  

\vs{10}
\end{abstract}
\end{titlepage}

\section{Introduction}

Non-perturbative analyses in 
quantum field theories and string theories 
have been recognized to be necessary 
to solve important problems which 
cannot be solved in the frameworks of perturbative theories. 
In some cases, 
degenerate vacua that are found to be stable 
in perturbative analyses turn out to be false vacua 
as a result of non-perturbative effects. 
Recently, there has been much progress in 
supersymmetric gauge field theories in 
four dimensions~\cite{SUSY-QCD}. 

Two-dimensional nonlinear sigma models have attracted interest
because of their similarity to four-dimensional gauge field theories, 
such as mass gaps, asymptotic freedom, instantons and so on 
(see, e.g., Refs.~\cite{Polyakov,Coleman,Abdalla,Hi} for a review). 
For this reason it is interesting to investigate 
non-perturbative effects in  
two-dimensional nonlinear sigma models.  
If we reformulate nonlinear sigma models 
by using auxiliary fields, 
we can investigate non-perturbative effects easily 
with large-$N$ methods. 
In the $O(N)$ model, we can find a mass gap, 
in contrast to perturbative analyses. 
In the ${\bf C}P^{N-1}$ model, there is a mass gap, 
a gauge boson is dynamically generated,  
and confinement due to this generated gauge boson 
occurs~\cite{DLD}.

The ${\cal N}=1$ supersymmetric $O(N)$ model consists of 
the bosonic $O(N)$ model and the Gross-Neveu model.  
Since the Gross-Neveu model illustrates 
dynamical chiral symmetry breaking, 
the ${\cal N}=1$ supersymmetric $O(N)$ model also has 
this property~\cite{WiAl}. 
In principle, bosonic and ${\cal N}=1$ supersymmetric 
nonlinear sigma models on an arbitrary coset space $G/H$ 
can be formulated using auxiliary fields 
to study non-perturbative effects. 

What about ${\cal N}=2$ supersymmetric 
nonlinear sigma models in two dimensions? 
These models may possess similarities to four-dimensional 
${\cal N}=2$ QCD.  
However, only a few models have been investigated to this time, 
since there is no auxiliary field formulation of 
${\cal N}=2$ supersymmetric nonlinear sigma models, 
except for the ${\bf C}P^{N-1}$ model 
and the Grassmannian model~\cite{Wi,DDL,Ao}. 
One of difficulties in such an investigation is the fact that 
target manifolds of ${\cal N}=2$ 
supersymmetric nonlinear sigma models must be 
K\"{a}hler manifolds~\cite{Zu}.
However, two of the present authors have recently given 
an auxiliary field formulation of four-dimensional ${\cal N}=1$ 
supersymmetric nonlinear sigma models 
(which are equivalent to two-dimensional  
${\cal N}=2$ supersymmetric nonlinear sigma models
through dimensional reduction),  
whose target spaces are the Hermitian symmetric spaces 
summarized in Table~\ref{table-HSS}~\cite{HN1,HN2}.
(For a review, see Ref.~\cite{HN3}.) 
\begin{table}[htbp]
\begin{center}
\begin{tabular}{c|c|c} \hline\hline
	Type & $G/H$ & ${\rm dim}_{\bf C} (G/H)$ \\ \hline
	AIII$_1$ & ${\bf C}P^{N-1} = SU(N)/SU(N-1) \times U(1)$ &
	$N-1$ \\
	AIII$_2$ & $G_{N,M} ({\bf C}) = U(N) / U(N-M) \times U(M)$ &
	$M(N-M)$ \\
	BDI & $Q^{N-2} ({\bf C}) = SO(N) / SO(N-2) \times U(1)$ & $N-2$ \\
	CI & $Sp (N) / U(N)$ & $\half N (N+1)$ \\
	DIII & $SO(2N) / U(N)$ & $\half N (N-1)$ \\
	EIII & $E_6 / SO(10) \times U(1)$ & 16 \\
	EVII & $E_7 / E_6 \times U(1)$ & 27 \\ \hline
\end{tabular}
\caption{Hermitian symmetric spaces.} \label{table-HSS}
\end{center}
{\footnotesize 
The left column gives the classification by Cartan.
The first three manifolds, 
${\bf C}P^{N-1}$, $G_{N,M}({\bf C})$ and $Q^{N-2}({\bf C})$,  
are called a (complex) projective space, 
a (complex) Grassmann manifold,  
and a (complex) quadric surface, respectively.
} 
\end{table}
The Hermitian symmetric spaces, which include the ${\bf C}P^{N-1}$ and 
the Grassmann manifold, 
are in a familiar class of K\"{a}hler coset spaces $G/H$, 
whose (nonlinear) K\"{a}hler potentials can be constructed 
by supersymmetric nonlinear realization methods~\cite{IKK}. 
As mentioned above, non-perturbative effects of 
the ${\cal N}=2$ supersymmetric ${\bf C}P^{N-1}$ model 
have been studied in a large number of works 
(see Refs.~\cite{Wi,DDL} and papers that cite them), 
since there already exists the auxiliary field formulation of 
${\bf C}P^{N-1}$. 
The other models have not been studied yet. 
The investigation of new models should provide 
deeper knowledge about non-perturbative effects 
of quantum field theories. 
The purpose of this paper is to investigate 
non-perturbative effects of one of the new models, 
the quadric surface 
$Q^{N-2}({\bf C})=SO(N)/SO(N-2) \times U(1)$, 
by using auxiliary field methods and large-$N$ methods 
(the leading order of the $1/N$ expansion). 
We call this model simply the ``$Q^N$ model'' in this paper. 

We introduce auxiliary vector and chiral superfields 
to reformulate the nonlinear sigma model. 
If we integrate out auxiliary superfields, 
we obtain the original nonlinear sigma model again. 
The integration over auxiliary vector and chiral superfields 
gives D-term and F-term constraints, respectively. 
If we set all of the auxiliary {\it chiral} superfields to zero, 
we obtain the ${\bf C}P^{N-1}$ model. 
If we set the auxiliary {\it vector} superfield to zero, 
we obtain the 
the non-compact ${\cal N}=2$ supersymmetric $O(N)$ sigma model,  
which is a generalization of bosonic and ${\cal N}=1$ 
supersymmetric $O(N)$ models~\cite{Ni}.
By integrating out the dynamical fields, 
and calculating the effective action and the effective potential, 
we investigate non-perturbatively stable vacua of 
the $Q^N$ model.
We find that there exist two stable vacua: 
one is the vacuum known in the ${\cal N}=2$ 
supersymmetric ${\bf C}P^{N-1}$ model, 
and the other is a new type of vacuum which 
is found here for the first time. 

We find that, in both phases, 
there is mass gap, and auxiliary superfields become dynamical 
as bound states of original dynamical fields for large $N$. 
In particular, a gauge boson is dynamically generated as 
in the ${\bf C}P^{N-1}$ model.  
Moreover, we show that both phases are asymptotically free 
by calculating the beta function.   
One of the key points which we should elucidate in both vacua is 
the disappearance of massless Nambu-Goldstone bosons. 
In two dimensions, the existence of
massless Nambu-Goldstone bosons is forbidden by 
Coleman's theorem~\cite{Co}. 
We can avoid this problem owing to several mechanisms,  
supersymmetry, the Schwinger mechanism, 
and the Higgs mechanism.

\vs{5}
This paper is organized as follows. 
In section \ref{formalism} we formulate 
the $Q^N$ model with auxiliary superfields 
using the notation of four-dimensional ${\cal N}=1$ supersymmetry. 
We perform dimensional reduction to two dimensions and   
summarize the symmetries of the two-dimensional Lagrangian.  
We calculate the effective potential and 
find two kinds of non-perturbatively stable vacua in section 3.    
In section \ref{Schwinger-phase} we discuss 
one of the stable vacua, which we call the Schwinger phase.   
In section \ref{Higgs-phase} 
we investigate the other stable vacuum,  
which is a new kind of vacuum, the Higgs phase. 
We devote section \ref{conclusion} to conclusions and discussion.
We summarize the notation of ${\cal N}=1$ supersymmetry in four 
dimensions in Appendix A. 
Appendix B describes the dimensional reduction 
to ${\cal N}=2$ supersymmetry in two dimensions.

\section{ Auxiliary Field Formulation of the $\mbf{Q^N}$ Model} 
\label{formalism}
In this section we formulate the $Q^N$ model 
with auxiliary superfields 
in four-dimensional ${\cal N}=1$ supersymmetry notation. 
We then perform the dimensional reduction to two dimensions 
following the prescription given in 
Appendix~\ref{dimensional-reduction}. 
We discuss symmetries of this Lagrangian 
in the second subsection. 

\subsection{Lagrangian in four dimensions 
and reduction to two dimensions}
First, we give the auxiliary field formulation of  
the $Q^N$ model in ${\cal N}=1$ 
four-dimensional notation~\cite{HN1,HN2}. 
Let $\Phi_i(x,\theta,\bar{\theta})$ ($i=1,\cdots, N$) be 
dynamical chiral superfields belonging to 
the $SO(N)$ vector representation. 
Then, the Lagrangian can be constructed  
by introducing auxiliary superfields as 
\begin{align}
  {\cal L}_{\rm linear} \ &= \ \int \! d^4 \theta 
  \big( \Phi_i^{\dagger} \Phi_i^{} e^{2V} - c V \big) 
  + \Big( \int \! d^2 \theta \, \Phi_0^{} \Phi_i^2 
         + \mbox{ h.c. } \Big) \; , \label{Q-L}
\end{align}
where $V(x,\theta,\bar{\theta})$ is an auxiliary vector superfield  
and $\Phi_0(x,\theta,\bar{\theta})$ is 
an auxiliary chiral superfield belonging to an $SO(N)$ singlet.  
Here, summation over the index $i$ is implied.  
The constant $c$ is positive and real, 
and the term $c V$ is the so-called Fayet-Iliopoulous term. 
The last two terms constitute the superpotential.  
This model has four-dimensional ${\cal N}=1$ supersymmetry, 
global $SO(N)$ symmetry, and $U(1)$ gauge symmetry:   
\begin{align}
 &\Phi_i(x,\theta,\bar{\theta}) \to 
   e^{i\Lambda(x,\theta,\bar{\theta})}
   \Phi_i(x,\theta,\bar{\theta}), \hs{10}
 \Phi_0(x,\theta,\bar{\theta}) \to 
   e^{-2 i\Lambda(x,\theta,\bar{\theta})}
   \Phi_0(x,\theta,\bar{\theta}), \nonumber\\
 &e^{2V(x,\theta,\bar{\theta})} 
 \to e^{2V(x,\theta,\bar{\theta}) 
       - i\Lambda(x,\theta,\bar{\theta}) 
       + i\Lambda(x,\theta,\bar{\theta})^{\dagger}}. 
 \label{gauge-4dim.}
\end{align}
Here $\Lambda(x,\theta,\bar{\theta})$ 
is an arbitrary chiral superfield. 
The partition function of this model can be written as 
\begin{align}
Z \ &= \ \int \! {\cal D} \Phi_i^{} {\cal D} \Phi_i^{\dagger} {\cal D}
\Phi_0^{} {\cal D} \Phi_0^{\dagger} {\cal D} V \, \exp \Big( i \int \!
d^4 x \, {\cal L}_{\rm linear} \Big) \; .  \label{partition}
\end{align}

The integration over $V(x,\theta,\thb)$ gives a D-term constraint, 
and the K\"{a}hler potential becomes nonlinear:  
\begin{align}
 {\cal L}' =\int \! d^4 \theta \; 
 c\log (\Phi_i^{\dagger} \Phi_i^{}).
\end{align} 
This is the K\"{a}hler potential of the Fubini-Study metric of 
${\bf C}P^{N-1}$ in the homogeneous coordinates $\Phi_i$. 
(We can show that integration over 
$V$ is equivalent to the elimination of $V$ 
by its classical equation of motion~\cite{HN2}.)
On the other hand, integration over $\Phi_0(x,\theta,\thb)$ gives 
the F-term constraint
\begin{align}
 \Phi_i(x,\theta,\thb)^2  =0,  \label{F-con.}
\end{align} 
which is holomorphic. 
${\bf C}P^{N-1}$ with the constraint 
$\Phi_i^2=0$ (on homogeneous coordinates $\Phi_i$) is 
just the (complex) quadric surface $Q^{N-2}({\bf C})$~\cite{KN}.   
We thus obtain the supersymmetric nonlinear sigma model
on  $Q^{N-2}({\bf C})$ 
by integration over the auxiliary superfields 
$V$ and $\Phi_0$~\cite{HN1,HN2}: 
\begin{align}
  {\cal L}_{\rm nonlinear} 
  \ &= \ \int \! d^4 \theta \,  c\, \log \Big\{ 1
   + \varphi_a^{\dagger} \varphi_a^{} + \frac{1}{4} \big(
  \varphi_a^{\dagger} \big)^2 \big( \varphi_a^{} \big)^2 \Big\} \; , 
   \label{QN-nonlinear}
\end{align}
where the fields $\varphi_a (x,\theta,\thb)$ ($a=1,\cdots,N-2$) 
are {\it nonlinear} dynamical chiral superfields. 
Here, a solution of the F-term constraint (\ref{F-con.}) on 
$\Phi_i = (\varphi_a,\alpha,\beta)$ is given by 
\begin{equation}
  \alpha - i \beta = - {(\varphi_a)^2 \over \alpha + i\beta} 
  = -{(\varphi_a)^2 \over \sqrt{2}}, \label{constraint} 
\end{equation}
where we have chosen the specific gauge 
$\alpha+i\beta=\sqrt{2}$ by using the gauge degrees of freedom, 
represented by Eq.~(\ref{gauge-4dim.}).\footnote{
If we choose another gauge, 
the K\"{a}hler potential in that gauge and 
Eq.~(\ref{QN-nonlinear})
can be transformed into each other 
by the K\"{a}hler transformation 
$K(\varphi,\varphi^{\dagger}) \to K(\varphi,\varphi^{\dagger}) 
+ f(\varphi) + f^{\dagger}(\varphi^{\dagger})$. 
The metrics in the two gauges coincide. 
}
The nonlinear Lagrangian (\ref{QN-nonlinear}) coincides 
with that constructed using  
the supersymmetric nonlinear realization methods~\cite{DV,IKK,Ni}. 
The target manifold parametrized by 
(scalar components of) $\varphi_a$ is the quadric surface 
and is isomorphic to a Hermitian symmetric space: 
\begin{align}
  Q^{N-2} ({\bf C}) \ &\simeq \ 
  \frac{\dps SO(N)}{\dps SO(N-2) \times U(1)} \; .
\end{align} 
Although the $\varphi_a$ transform nonlinearly under $SO(N)$, 
they transform linearly 
under the isotropy group $SO(N-2) \times U(1)$, 
namely as an $SO(N-2)$ vector with an appropriate $U(1)$ charge. 

Let us construct the two-dimensional model using
the notation and reduction rules summarized in 
Appendix~\ref{dimensional-reduction}. 
The superfields and their component fields are 
\begin{align}
 &\Phi_i(x,\theta,\thb): \;\; (A_i(x),\psi_i(x),F_i(x)),\hs{10}
  \Phi_0(x,\theta,\thb): \;\; (A_0(x),\psi_0(x),F_0(x)),\nonumber\\
 &V(x,\theta,\thb): \;\; (M(x),N(x), \lambda(x), V_m(x), D(x)),
\end{align}
where the $\Phi_i(x,\theta,\thb)$ are dynamical chiral superfields, 
and $\Phi_0(x,\theta,\thb)$ and $V(x,\theta,\thb)$ 
are auxiliary chiral and vector superfields, respectively. 
Instead of choosing the gauge (\ref{constraint}) we have assumed the
{\it Wess-Zumino gauge} for $V$, suitable for the non-perturbative study.
The index $m \,(=0,1)$ labels the two-dimensional space-time coordinates  
$(x^0,x^1)$. 
The components fields $N(x)$ and $M(x)$ in $V$ 
are real scalar fields in two dimensions 
that originate from components of 
the gauge field $V_{\mu}(x)$ in four dimensions, 
$V^2(x)$ and $V^3(x)$, respectively:  
\begin{align*} 
 N(x) \equiv V^2(x) \ = \ - V_2(x) \ , \ \ \ 
 M(x) \equiv V^3(x) \ = \ - V_3(x) \; . 
\end{align*}
Furthermore, we redefine the gauginos $\lambda$, 
$\lambda^c$ and complex scalar fields $A_0^{}$, $A_0^*$ by  
\begin{subequations}
\begin{align}
 &\lambda \ \to \ \sqrt{2} i \lambda \; , \ \ \ \lambda^c \ \to \ -
 \sqrt{2} i \lambda^c \; , \\ 
 &A_0^{} \ \to \ \half A_0
 \; , \ \ \ A_0^* \ \to \ \half A_0^* \; ,
\end{align}
\end{subequations}
where $\lambda^c$ is the charge conjugate of $\lambda$:  
$\lambda^c = - \gamma^0 \ol{\lambda}^T$. 
We thus obtain the auxiliary field formulation 
of the two-dimensional $Q^N$ model 
in component fields, given by  
\begin{align}
 {\cal L} \ &= \ F_i^* F_i^{} + \del_m A_i^* \del^m A_i^{} + i
 \ol{\psi}_i \gamma^m \del_m \psi_i^{} \nonumber \\
 & \ \ \ \ + V_m \big[ i A_i^* \del^m A_i^{} - i \del^m A_i^* \cdot
 A_i^{} + \ol{\psi}_i \gamma^m \psi_i^{} \big] + M \big( \ol{\psi}_i
 \psi_i^{} \big) - N \big( \ol{\psi}_i i \gamma_3 \psi_i^{} \big)
 \nonumber \\
 & \ \ \ \ + A_i^{} \big( \ol{\lambda} \psi_i^c +
 \ol{\psi}_i \lambda^c \big) + A_i^* \big(
 \ol{\lambda^c} \psi_i^{} + \ol{\psi_i^c} \lambda \big)
 \nonumber \\
 & \ \ \ \ + ( D + V_m V^m - M^2 - N^2 ) A_i^* A_i^{} - \half c D
 \nonumber \\
 \ & \ \ \ \ + \big\{ F^{}_0 A^2_i + F^*_0 A^*_i{}^2 
 \big\} + \big\{ F^{}_i A^{}_i A^{}_0 + F^{*}_i A^{*}_i A^{*}_0
 \big\} \nonumber \\
 \ & \ \ \ \ - A_i^{} \big( \ol{\psi_0^c} \psi_i^{} + \ol{\psi_i^c}
 \psi_0^{} \big) - A_i^* \big( \ol{\psi}_0 \psi_i^c + \ol{\psi}_i
 \psi_0^c \big) - \half A_0^{} \ol{\psi_i^c} \psi_i^{} - \half A_0^*
 \ol{\psi}_i \psi_i^c \; , \label{Q-L-2D}
\end{align}
where $\psi_i^c$ is the charge conjugate of $\psi_i$.

We can eliminate the auxiliary fields for supersymmetry, 
$F_i(x)$ and $F_i^*(x)$, 
in the dynamical chiral superfields $\Phi_i$
by substituting their equations of motion,  
\begin{align}
 F_i^{}(x) \ &= \ - A_i^*(x) A_0^*(x) \, , \hs{10} 
 F_i^*(x) \ = \ - A_i^{}(x) A_0^{}(x)\, ,  \label{F-consistency}
\end{align}
back into the Lagrangian (\ref{Q-L-2D}):
\begin{align}
{\cal L} \ &= \del_m A_i^* \del^m A_i^{} + i
\ol{\psi}_i \gamma^m \del_m \psi_i^{} \nonumber \\
& \ \ \ \ + V_m \big[ i A_i^* \del^m A_i^{} - i \del^m A_i^* \cdot
A_i^{} + \ol{\psi}_i \gamma^m \psi_i^{} \big] + M \big( \ol{\psi}_i
\psi_i^{} \big) - N \big( \ol{\psi}_i i \gamma_3 \psi_i^{} \big)
\nonumber \\
& \ \ \ \ + A_i^{} \big( \ol{\lambda} \psi_i^c +
\ol{\psi}_i \lambda^c \big) + A_i^* \big(
\ol{\lambda^c} \psi_i^{} + \ol{\psi_i^c} \lambda \big)
\nonumber \\
& \ \ \ \ + ( D + V_m V^m - M^2 - N^2 ) A_i^* A_i^{} - \half c D
\nonumber \\
\ & \ \ \ \ + F^{}_0 A^2_i + F^*_0 A^*_i{}^2 
- A_0^* A_0^{} A_i^* A_i^{} \nonumber \\
\ & \ \ \ \ - A_i^{} \big( \ol{\psi_0^c} \psi_i^{} + \ol{\psi_i^c}
\psi_0^{} \big) - A_i^* \big( \ol{\psi}_0 \psi_i^c + \ol{\psi}_i
\psi_0^c \big) - \half A_0^{} \ol{\psi_i^c} \psi_i^{} - \half A_0^*
\ol{\psi}_i \psi_i^c \; . \label{Q-new-L}
\end{align}
Although we start from this Lagrangian (\ref{Q-new-L}) 
in the following sections, 
we discuss the symmetries of (\ref{Q-L-2D}) 
in the next subsection.   

Before discussing symmetries, 
we eliminate the remaining auxiliary fields.  
If we eliminate all auxiliary fields 
using their equations of motion, 
we obtain the nonlinear Lagrangian 
\begin{subequations}
\begin{align}
 {\cal L} = (D_m A_i)^* (D^m A_i) 
 + i \ol{\psi}_i \gamma^m D_m \psi_i 
 + {1\over 2c}\left[(\ol{\psi}_i\psi_i)^2 
     + (\ol{\psi}_i i \gamma_3 \psi_i)^2 
     + (\ol{\psi}_i^{}\psi_i^c)(\ol{\psi}_i^c \psi_i^{}) \right],
  \label{2-dim.nonlinear_lag}
\end{align}
with the constraints
\begin{align}
 A_i^* A_i^{} = {c\over 2}, \hs{10} 
 A_i^* \psi_i^{} = A_i \ol{\psi}_i^{} = 0, \hs{10} 
 A_i \psi_i^{} = A_i^* \ol{\psi}_i^{} = 0, \hs{10} 
 A_i^2 = A_i^{*2}  = 0. 
\end{align}
\end{subequations}
The first (and the second) equations are the same 
as those of the (supersymmetric) ${\bf C}P^N$ model,  
and the last two are the same as those of  
the ${\cal N}=1$ supersymmetric $O(N)$ model with 
``zero radius''.\footnote{
Since the $A_i(x)$ are complex scalar fields, 
$A_i^2 = 0$ does not represent a point, 
but a conifold~\cite{Ni}.
}  
In Eq.~(\ref{2-dim.nonlinear_lag}), 
$D_m$ is the covariant derivative defined by
\begin{subequations}
\begin{align}
 &D_m A_i = (\del_m - i V_m) A_i \; , \hs{10} 
   D_m \psi_i = (\del_m - i V_m) \psi_i \; , \\
 &V_m = {1 \over c} \left(i A^*_i 
     {\stackrel{\leftrightarrow}{\del}}_m A_i^{} 
     + \bar\psi_i \gamma_m \psi_i \right).
\end{align}
\end{subequations}
The nonlinear Lagrangian (\ref{2-dim.nonlinear_lag}) 
can be obtained from 
Eq.~(\ref{QN-nonlinear}) by eliminating auxiliary fields 
for supersymmetry in $\varphi_a(x,\theta,\thb)$ 
(and by dimensional reduction).

\subsection{ Symmetries }
In this subsection 
we consider symmetries of 
the two-dimensional Lagrangian (\ref{Q-L-2D}),  
before eliminating the auxiliary fields $F_i$ and $F_i^*$. 
The Lagrangian (\ref{Q-L-2D}) has three types of $U(1)$ symmetries, 
as described below. 

\begin{enumerate}

\item The local (gauged) $U(1)$ symmetry: 

This is a local phase transformation on superfields  
under which the Grassmannian coordinates $\theta$ are invariant. 
It is given by  
\begin{align}
 \Phi_i (x, \theta, \ol{\theta}) \ &\to 
 \ e^{i \alpha (x,\theta,\thb)} \Phi_i (x,\theta, \ol{\theta}) \; ,
 \ \ \ \Phi_0 (x, \theta, \ol{\theta}) \ \to \ 
 e^{-2 i \alpha (x,\theta,\thb)}
\Phi_0 (x, \theta, \ol{\theta}) \; ,
\end{align}
where $\alpha (x,\theta,\thb)$ is 
an arbitrary chiral superfield gauge parameter. 

\item The global $U(1)$ symmetry: 

This is the global phase transformation 
on the Grassmannian parameters $\theta$, namely the $R$ symmetry, 
\begin{subequations}
\begin{align}
 \Phi_i (x, \theta, \ol{\theta}) \ &\to \ 
   \Phi_i (x, e^{i \alpha} \theta,
    e^{-i \alpha} \ol{\theta}) \; , \ \ \ 
 \Phi_0 (x, \theta, \ol{\theta}) \ \to \ e^{2 i \alpha} 
    \Phi_0 (x, e^{i\alpha} \theta, e^{-i \alpha} \ol{\theta})
\; , \\
\lambda (x) \ &\to \ e^{i \alpha} \lambda (x) \; .
\end{align}
\end{subequations}
The origin of this symmetry is   
the $R$ symmetry in four dimensions.

\item The global chiral $U(1)$ symmetry: 

This $U(1)$ symmetry is another 
$R$ symmetry whose origin is the rotation in 
the $(x^2,x^3)$-plane in four dimensions. 
In two dimensions, this becomes a chiral symmetry, given by 
\begin{subequations}\label{chiral-sym.}
\begin{align}
 &\Phi_i (x, \theta, \ol{\theta}) \ \to \ 
   \Phi_i (x, e^{i \gamma_3\alpha} \theta, 
           \ol{\theta} e^{i \gamma_3 \alpha}) \; , \ \ \ 
\Phi_0 (x, \theta, \ol{\theta}) \ \to \ 
\Phi_0 (x, e^{i \gamma_3 \alpha} \theta, 
        \ol{\theta} e^{i \gamma_3 \alpha}) \; , \\
&\lambda (x) \ \to \ e^{i \gamma_3 \alpha} \lambda (x) \; , \ \ 
M (x) - i \gamma_3 N (x) \ \to \ e^{-2 i \gamma_3 \alpha} 
   \big( M(x) - i\gamma_3 N (x) \big) \; . 
\end{align}
\end{subequations}

\end{enumerate}
We list the charges of the superfields 
$\Phi_i$, $\Phi_0$ and $V$,  
their component fields, and the Grassmannian variables $\theta$  
under transformations of these symmetries 
in Table~\ref{symmetry-charge}.
\begin{table}[htbp]
\begin{center}
\begin{tabular}{c||cccc|cccc|ccccc|c}
	symmetries & $\Phi_i$ & $A_i$ & $\psi_i$ & $F_i$ & $\Phi_0$ &
	$A_0$ & $\psi_0$ & $F_0$ & $V$ & $V_m$ & $M - i \gamma_3 N$ &
	$\lambda$ & $D$ & $\theta$ \\ \hline\hline
	local $U(1)$ & $1$ & $1$ & $1$ & $1$ & $-2$ & $-2$ & $-2$ & $-2$ &
	$0$ & $0$ & $0$ & $0$ & $0$ & $0$ \\
	global $U(1)$ & $0$ & $0$ & $-1$ & $-2$ & $2$ & $2$ & $1$ & $0$ &
	$0$ & $0$ & $0$ & $1$ & $0$ & $1$ \\
	global chiral $U(1)$ & $0$ & $0$ & $1$ & $0$ & $0$ & $0$ & $1$
	& $0$ & $0$ & $0$ & $-2$ & $1$ & $0$ & $1$ \\ \hline
	global + local $U(1)$ & $1$ & $1$ & $0$ & $-1$ & $0$ & $0$ &
	$-1$ & $-2$ & $0$ & $0$ & $0$ & $1$ & $0$ & $1$ 
\end{tabular}
\caption{$U(1)$ symmetries and their charges.} 
{\footnotesize  
The last line gives the mixed $U(1)$ symmetry of 
the global $U(1)$ and the local $U(1)$, 
which we consider below.
}
\label{symmetry-charge}
\end{center}
\end{table}

\section{ Effective Potential and Vacua } \label{effpot}
In this section, we calculate the effective potential by 
integrating out all of the dynamical fields 
$A_i(x)$ and $\psi_i(x)$. From the variations of 
the effective potential
with respect to the vacuum expectation values, 
we obtain gap equations.  
By solving them we find two kinds of stable vacua.  

\subsection{ Effective potential and vacuum conditions }
We start from the Lagrangian (\ref{Q-new-L}). 
We would like to find non-perturbative vacua 
by analyzing the leading order of the $1/N$ expansion. 
At the leading order in this expansion,
we can neglect quantum fluctuations of auxiliary fields. 
We set the vacuum expectation values 
of auxiliary fields as 
\begin{align}
 A_0^{} (x) \ &= \ \phi_0^{} \,, \hs{10} 
 A_0^* (x) \ = \ \phi_0^* \,, \hs{10} 
 F_0^{} (x) \ = \ F_c^{} \,, \hs{10}
 F_0^* (x) \ = \ F_c^* \, ,\nonumber \\
 \psi_0^{} (x) \ &= \ \ol{\psi}_0 (x) \ = \ \lambda (x) \ = \
 \ol{\lambda} (x) \ = \ 0 \; , \nonumber \\
 M (x) \ &= \ M_c \, ,\hs{10} N (x) \ = \ N_c \,,\hs{10} 
 D (x) \ = \ D_c \, ,\hs{10} V_m (x) \ = \ 0 \,.\label{c-fields}
\end{align}
Moreover, we decompose the dynamical scalar fields $A_i (x)$
into sums of the classical constant fields $\phi_i$ 
and the fluctuating quantum fields $A_i'(x)$ around them,  
with constraints
\begin{align}
\int \! d^2 x \, A_i' (x) \ = \ 0 \; .
\end{align}
Since the vacuum expectation values of 
the dynamical fermionic fields $\psi_i (x)$ are all zero, 
we express these fluctuating fields also by $\psi_i (x)$.


Let us calculate the effective potential. 
We substitute the constant fields
of the auxiliary fields (\ref{c-fields}) into Eq.~(\ref{Q-new-L}). 
By integrating out the fluctuating dynamical fields 
$A_i'(x)$ and $\psi_i(x)$ 
in the partition function (\ref{partition}), 
\begin{align}
 & Z \ = \ \int \! {\cal D} \Phi_0^{} {\cal D} \Phi_0^{\dagger} 
 {\cal D}V \, \exp \big( i S_{\rm eff} \big) \; , 
\end{align}
we can calculate the effective action $S_{\rm eff}$, given by 
\begin{align}
 S_{\rm eff} \ &= \ \frac{i N}{2} {\rm Tr} \log \det 
  \big[ D_c^{-1}\big]  - \frac{i N}{2} 
  {\rm Tr} \log \det \big[ S_c^{-1} \big] + \int
  \! d^2 x {\cal L}_0 \; . \label{Q-effective-action}
\end{align}
Here we have defined 
\begin{subequations}
\begin{align}
D_c^{-1} \ &= \ \left(
	\begin{array}{cc}
	\del^2 + \phi_0^* \phi_0^{} - D_c + M_c^2 + N_c^2 & -2 F_c^* \\
	- 2 F_c^{} & \del^2 + \phi_0^* \phi_0^{} - D_c + M_c^2 +
	N_c^2
	\end{array} \right) \; , \\
S_c^{-1} \ &= \ \left(
	\begin{array}{cc}
	i \gamma^m \del_m + M_c \cdot {\bf 1} - i \gamma_3 N_c  & -
	\phi_0^* \cdot {\bf 1} \\
	- \phi_0^{} \cdot {\bf 1} & i \gamma^m \del_m + M_c \cdot
{\bf 1} + i \gamma_3 N_c
	\end{array} \right) \; , \\
{\cal L}_0 \ &= \ F_c^{} \phi_i^2 + F_c^* \phi_i^*{}^2 - \phi_0^*
\phi_0^{} \phi_i^* \phi_i^{} + ( D_c - M_c^2 - N_c^2 ) \phi_i^*
\phi_i^{} - \frac{N}{g^2} D_c \; .
\end{align}
\end{subequations}
We also have set the Fayet-Iliopoulous constant $c$ as
\begin{align}
  c \ &= \ \frac{2 N}{g^2} \; ,
\end{align}
by using the coupling constant $g$ and the numbers of 
dynamical fields $A_i$ and $\psi_i$, $N$. 
In this definition, all terms in the
Lagrangian (\ref{Q-new-L}) become of order $N$. 
The effective potential $V_{\rm eff}$ can be calculated from  
the definition 
\begin{align}
 S_{\rm eff} \big|_{\mbox{\footnotesize constant fields}} \ = \ -
 V_{\rm eff} \int \! d^2 x \; 
\end{align}
to give 
\begin{align}
 V_{\rm eff} \ &= \ \frac{N}{2} \int \! \frac{d^2 k}{(2\pi)^2 i} \log
 \Big[ (-k^2 + X^2 + Y^2 - D_c )^2 - 4 F_c^* F_c^{} \Big] \nonumber \\
 & \ \ \ \ - \frac{N}{2} \int \! \frac{d^2 k}{(2\pi)^2 i} \log \Big[
 ( -k^2 + X^2 + Y^2 )^2 - 4 X^2 Y^2 \Big] \nonumber \\
 & \ \ \ \ - F_c^{} \phi_i^2 - F_c^* \phi_i^*{}^2 + ( X^2 + Y^2 - D_c )
 \phi_i^* \phi_i^{} + \frac{N}{g^2} D_c \;. 
 \label{Q-effective-potential}
\end{align}
Here we have defined $X^2$ and $Y^2$ by 
\begin{align}
  Y^2 \ &\equiv \ M_c^2 + N_c^2 \; , \ \ \
  X^2 \ \equiv \phi_0^* \phi_0^{} \; ,
\end{align}
which play the roles of order parameters of the two kinds of vacua,  
as seen in the following sections. 


We can find non-perturbative vacua as 
the minimum points of the effective potential 
(\ref{Q-effective-potential}).  
By variations about all of the constant fields 
$\phi_i^{}$, $\phi_i^*$, $X$, $Y$, $F_c^{}$, $F_c^*$ and $D_c^{}$,  
we obtain the conditions for vacua, 
which are called the {\it gap equations}:  
\begin{subequations}
\begin{align}
0 \ &= \ \frac{N}{g^2} - \phi_i^* \phi_i^{} - N \int \!
\frac{d^2 k}{(2 \pi)^2 i} \frac{ -k^2 + X^2 + Y^2 - D_c }{ (
-k^2 + X^2 + Y^2 - D_c )^2 - 4 F_c^* F_c^{} } \; ,
\label{Q-condition-1} \\
0 \ &= \ - \phi_i^2 - 2 N \int \! \frac{d^2 k}{(2 \pi)^2i} 
  \frac{F_c^*}{ (- k^2 + X^2 + Y^2 - D_c )^2 - 4 F_c^*F_c^{}}\;, 
    \label{Q-condition-2} \\
0 \ &= \ - \phi_i^*{}^2 - 2 N \int \! \frac{d^2 k}{(2 \pi)^2 i} 
  \frac{F_c^{}}{ (- k^2 + X^2 + Y^2 - D_c )^2 - 4 F_c^* F_c^{} }\;, 
    \label{Q-condition-3} \\
0 \ &= \ \phi_i^{} \big\{ 4 F_c^* F_c^{} 
  - ( X^2 + Y^2 - D_c)^2 \big\}\;, \label{Q-condition-4} \\
0 \ &= \ \phi_i^* \big\{ 4 F_c^* F_c^{} 
   - ( X^2 + Y^2 - D_c)^2\big\} \; , \label{Q-condition-5} \\
0 \ &= \ 2 X \left\{ \frac{N}{g^2} - N \int \! 
  \frac{d^2 k}{(2 \pi)^2 i} \frac{-k^2 + X^2 - Y^2}
  {(-k^2 + X^2 + Y^2)^2 - 4 X^2 Y^2} \right\} \; ,
  \label{Q-condition-6} \\
0 \ &= \ 2 Y \left\{ \frac{N}{g^2} - N \int \! 
  \frac{d^2 k}{(2 \pi)^2 i} \frac{-k^2 - X^2 + Y^2}
  {(-k^2 + X^2 + Y^2)^2 - 4 X^2 Y^2} \right\} \;.
    \label{Q-condition-7}
\end{align}
\end{subequations}
We can find non-perturbatively stable vacua from these equations. 

\subsection{ Supersymmetric vacua }
Since we would like to find the supersymmetric vacua,
we assume the conditions
\begin{align}
  F_c^{} \ &= \ F_c^* \ = \ D_c^{} \ = \ 0 \; . 
  \label{super-conditions}
\end{align}
If we can find stable supersymmetric vacua, we need not search
for non-supersymmetric (supersymmetry broken) vacua,  
since non-supersymmetric vacua are unstable 
if supersymmetric vacua exist. 
Only when we cannot find any stable vacua preserving supersymmetry 
under these conditions 
should we search for non-supersymmetric stable vacua. 
Substitution of Eq.~(\ref{super-conditions})  
into the gap equations, 
Eqs.~(\ref{Q-condition-1})-(\ref{Q-condition-7}),  
gives 
\begin{subequations}
\begin{align}
& \phi_i^{} \big[ X^2 + Y^2 \big] \ = \ 0 \; , \ \ \ \phi_i^* \big[
X^2 + Y^2 \big] \ = \ 0 \; , \ \ \ \phi_i^2 \ = \ \phi_i^*{}^2 \ = \ 0
 \; , \label{Q-condition-8} \\
& \frac{N}{g^2} \ = \ \phi_i^* \phi_i^{} + N \int \! \frac{d^2 k}{(2
\pi)^2 i} \frac{1}{-k^2 + X^2 + Y^2} \; , \label{Q-condition-9} \\
& 0 \ = \ 2 X \left\{ \frac{N}{g^2} - N \int \! \frac{d^2 k}{(2 \pi)^2
i} \frac{-k^2 + X^2 - Y^2}{(-k^2 + X^2 + Y^2)^2 - 4 X^2 Y^2} \right\}
\; , \label{Q-condition-10} \\
& 0 \ = \ 2 Y \left\{ \frac{N}{g^2} - N \int \! \frac{d^2 k}{(2 \pi)^2
i} \frac{-k^2 - X^2 + Y^2}{(-k^2 + X^2 + Y^2)^2 - 4 X^2 Y^2} \right\}
\; . \label{Q-condition-11}
\end{align}
\end{subequations}
The last two equations show that at least 
$X$ or $Y$ must be zero; 
these equations are inconsistent with 
$X \neq 0$ and $Y\neq 0$ holding simultaneously.   
We thus find 
two kinds of consistent vacua, $X=0$ and $Y=0$.  
We discuss these below. 

\begin{enumerate}

\item 
In the $X=0$ vacuum,  
Eq.~(\ref{Q-condition-10}) is trivially satisfied,   
and the other equations become as follows:  
\begin{align}
\mbox{(\ref{Q-condition-8})} \ &: \ \phi_i^{} Y^2 \ = \ \phi_i^* Y^2
\ = \ \phi_i^2 \ = \ \phi_i^*{}^2 \ = \ 0 \; , \label{part1-1} \\
\mbox{(\ref{Q-condition-9})} \ &: \ \frac{N}{g^2} \ = \ \phi_i^*
\phi_i^{} + N \int \! \frac{d^2 k}{(2 \pi)^2 i} \frac{1}{-k^2 + Y^2} \; ,
\label{part1-2} \\
\mbox{(\ref{Q-condition-11})} \ &: \ 0 \ = \ 2 Y \left\{ \frac{N}{g^2} - N
\int \! \frac{d^2 k}{(2 \pi)^2 i} \frac{1}{-k^2 + Y^2} \right\} \;
. \label{part1-3}
\end{align}
If $Y^2 = 0$, Eq.~(\ref{part1-2}) would contain 
an infrared divergence. 
Since this divergence is singular, 
we cannot renormalize it. 
We thus conclude that $Y^2 \neq 0$. 
Under this condition, 
we obtain the final form of the gap equations, 
\begin{align}
\phi_i \ &= \ \phi_i^* \ = \ 0 \; , \ \ \ \frac{N}{g^2} \ = \ N \int
\! \frac{d^2 k}{(2 \pi)^2 i} \frac{1}{-k^2 + Y^2} \; . 
  \label{gap_eq_Sh.}
\end{align}
The second equation is the same as the gap equation 
of the bosonic $O(N)$ model 
for zero vacuum expectation values. 
The value of the effective potential under these conditions is zero:  
\begin{align}
  V_{\rm eff} \ &= \ 0 \; . 
\end{align}
This vacuum has some interesting features. 
First, it is supersymmetric. 
Second, the vacuum expectation values $\phi_i$ of
the dynamical scalar fields $A_i(x)$, 
belonging to an $SO(N)$ vector, are all zero.  
In perturbation theories, 
their values are nonzero and $SO(N)$ symmetry is broken.  
Contrastingly in non-perturbative vacua, 
we obtain zero vacuum expectation values 
of $A_i (x)$ and find that $SO(N)$ symmetry is restored. 
Third, all dynamical fields acquire masses 
$m = Y$ in order to avoid the infrared divergence. 
In particular, Dirac fermions $\psi_i(x)$ acquire 
{\it Dirac} mass terms. 
We call this vacuum the ``{\bf Schwinger phase}'', 
since the gauge field becomes massive as a result of 
the Schwinger mechanism, 
as shown in the next section~\cite{schwinger}.
This vacuum is the same as 
that of the ${\cal N}=2$ supersymmetric ${\bf C}P^{N-1}$ model.

\vs{5}

\item 
In the $Y=0$ vacuum,  
Eq.~(\ref{Q-condition-11}) is trivially satisfied,  
and the other conditions become 
\begin{align}
 \mbox{(\ref{Q-condition-8})} \ &: \ 
  \phi_i^{} X^2 \ = \ \phi_i^* X^2\ 
  = \ \phi_i^2 \ = \ \phi_i^*{}^2 \ = \ 0 \; , \label{part2-1} \\
\mbox{(\ref{Q-condition-9})} \ &: \ \frac{N}{g^2} \ = \ \phi_i^*
\phi_i^{} + N \int \! \frac{d^2 k}{(2 \pi)^2 i} \frac{1}{-k^2 + X^2} \; ,
\label{part2-2} \\
\mbox{(\ref{Q-condition-10})} \ &: \ 0 \ = \ 2 X \left\{ \frac{N}{g^2} - N
\int \! \frac{d^2 k}{(2 \pi)^2 i} \frac{1}{-k^2 + X^2} \right\} \;
. \label{part2-3}
\end{align}
If $X^2 = 0$, Eq.~(\ref{part2-2}) would contain 
an infrared divergence. 
Hence, in order to avoid this divergence, 
we should have $X^2 \neq 0$. 
We thus obtain the final form of the gap equations,  
\begin{align}
 \phi_i \ &= \ \phi_i^* \ = \ 0 \; , \ \ \ 
 \frac{N}{g^2} \ = \ N \int\! \frac{d^2 k}{(2 \pi)^2 i} 
                           \frac{1}{-k^2 + X^2} \; , 
   \label{gap_eq_Hi.}
\end{align}
which are the same as Eq.~(\ref{gap_eq_Sh.}) 
if we replace $X^2$ by $Y^2$.
The value of the effective potential is again zero: 
\begin{align}
V_{\rm eff} \ &= \ 0 \; . 
\end{align}
This vacuum has also some interesting features. 
It is supersymmetric and $SO(N)$ symmetric, 
and there are mass gaps about all fields. 
In this vacuum, Dirac spinors $\psi_i(x)$ obtain 
{\it Majorana} mass terms, in contrast to the Schwinger phase.  
We call this vacuum the ``{\bf Higgs phase}'', 
since a gauge boson acquires mass through the Higgs mechanism. 
This vacuum had until this time not been seen in 
two-dimensional nonlinear sigma models.

\end{enumerate}

We thus have found two stable vacua, 
the Schwinger phase, in which $Y\neq 0$ and $X=0$, 
and the Higgs phase, in which $X\neq 0$ and $Y=0$. 
The final form of the gap equations,  
Eqs.~(\ref{gap_eq_Sh.}) and (\ref{gap_eq_Hi.}), 
is the same in the two phases 
if we replace $X$ by $Y$ and vice versa.
The Schwinger phase and the Higgs phase in this model  
are similar to the Coulomb and Higgs branches 
in ${\cal N}=2$ supersymmetric QCD in four dimensions, 
where scalar components of vector-multiplets 
and hyper-multiplets acquire vacuum expectation values, respectively. 
In the following two sections, 
we calculate two-point functions and $\beta$ functions 
in these two vacua. 

\section{ Schwinger Phase } \label{Schwinger-phase}
In this section we investigate the Schwinger phase, 
which is well known 
as the non-perturbative vacuum of 
the ${\cal N}=2$ supersymmetric 
${\bf C}P^{N-1}$ model~\cite{Wi,DDL}. 
In this phase, components of the dynamical superfields, 
belonging to $SO(N)$ vectors, 
and components of the auxiliary superfields 
acquire the non-zero masses $m = |Y|$ and $m = |2Y|$, respectively.  
We find that all the auxiliary superfields become 
dynamical as bound states of the original dynamical fields.
We also calculate propagators and the $\beta$ function, 
and find that this phase is asymptotically free.

\subsection{ Two-point functions }
In this phase, $Y^2=M_c^2 +N_c^2$ is the only non-zero 
vacuum expectation value. 
By using the chiral symmetry Eq.~(\ref{chiral-sym.}), 
this vacuum expectation value can be rotated to $M_c$. 
Then vacuum expectation values are given by  
\begin{subequations}\label{VEVs-in-Sh.}
\begin{align}
 & \phi_0^{} \ = \ \phi_0^* \ = \ F_c^{} \ = \ F_c^* \ = \ 0 \; , \\
 & N_c \ = \ D_c \ = \ \langle \lambda \rangle \ 
         = \ \langle \psi_0 \rangle \ 
         = \ \langle V_m \rangle \ = \ 0 \; , \\
 & M_c \ = \ - Y \ \neq \ 0 \; . 
\end{align}
\end{subequations}
We define $M'$ and $N'$ 
as the quantum fluctuations of $M$ and $N$ 
around the above vacuum expectation values. 
For quantum fluctuations of the remaining fields, 
the same letters are used, 
since their vacuum expectation values are all zero.
By the vacuum expectation values (\ref{VEVs-in-Sh.}), 
the chiral $U(1)$ symmetry is spontaneously broken, 
as seen in Table~\ref{symmetry-charge}. 
Then $N'$ is a Nambu-Goldstone boson of this breaking, 
and is massless.  
(As shown below, this massless boson disappears from 
the physical spectrum, because it is absorbed by a gauge boson.) 
Under Eq.~(\ref{VEVs-in-Sh.}), 
the Lagrangian (\ref{Q-new-L}) becomes 
\begin{align}
{\cal L} \ &= \ - A_i^* \big[ \del^2 + Y^2 \big] A_i + \half \Big( \
	\ol{\psi}_i \; , \;
	\ol{\psi_i^c} \ \Big) \left(
	\begin{array}{cc}
	i \gamma^m \del_m - Y & 0 \\
	0 & i \gamma^m \del_m - Y
	\end{array} \right) \left(
	\begin{array}{c}
	\psi_i \\
	\psi_i^c
	\end{array} \right) \nonumber \\
& \ \ \ \ + V_m \Big( i A_i^* \del^m A_i^{} - \del^m A_i^* \cdot
	A_i^{} + \ol{\psi}_i \gamma^m \psi_i \Big) + V_m V^m
	A_i^* A_i^{} \nonumber \\
& \ \ \ \ + A_i^{} \Big( \ol{\lambda}
	\psi_i^c + \ol{\psi}_i \lambda^c \Big) +
	A_i^* \Big( \ol{\lambda^c} \psi_i^{} + \ol{\psi_i^c} \lambda
	\Big) - A_i^{} \Big( \ol{\psi_0^c} \psi_i^{} +
	\ol{\psi_i^c} \psi_0^{} \Big) - A_i^* \Big( \ol{\psi}_0
	\psi_i^c + \ol{\psi}_i \psi_0^c \Big) \nonumber \\
& \ \ \ \ - \Big( - 2 Y M' + M'{}^2 + N'{}^2 \Big) A_i^* A_i^{} + M'
	\Big( \ol{\psi}_i \psi_i^{} \Big) - N' \Big( \ol{\psi}_i i
	\gamma_3
	\psi_i^{} \Big) \nonumber \\
& \ \ \ \ - A_0^* A_0^{} A_i^* A_i^{} - \half A_0^{} \ol{\psi_i^c}
	\psi_i^{} - \half A_0^* \ol{\psi}_i \psi_i^c \nonumber \\
& \ \ \ \ + D A_i^* A_i^{} - \frac{N}{g^2} D + F_0^* A_i^2 + F_0^*
	A_i^*{}^2 \; . \label{Q-schwinger-L}
\end{align}
~From this equation, 
we find that the original dynamical spinors $\psi_i$ acquire 
the Dirac mass terms $Y \ol{\psi}_i \psi_i^{}$. 
The auxiliary spinor $\psi_0$ also acquires  
the Dirac mass term, as seen in Table \ref{chiral-DS}, below.

We now expand the effective action to calculate two-point functions. 
We define two-point functions as coefficients of 
quadratic terms in an expansion of the effective action by 
\begin{align}
  S_{\rm eff} \ &= \ \int \! \frac{d^2 p}{(2 \pi)^2} \sum_{i,j} 
  \wt{\cal F}_i (-p) \Pi_{{\cal F}_i {\cal G}_j} (p) 
  \wt{\cal G}_j (p) + \cdots \; , 
\end{align}
where $\wt{\cal F}_i(p)$ and $\wt{\cal G}_i (p)$ are arbitrary fields
in the momentum representation, and the coefficients 
$\Pi_{{\cal F}_i {\cal G}_j}(p)$ are their two-point functions. 
We thus obtain all of the two-point functions in this phase 
as listed in Tables~\ref{chiral-DS} and \ref{vector-DS}. 
In these tables, for simplicity, 
we have omitted multiplication 
by the factor $R(p^2)$ defined by 
\begin{align}
R(p^2) \ &= \ \frac{N}{2 \pi} \int_0^1 \! dx \ 
  \frac{1}{Y^2 - x(1-x) p^2}\; . \label{R-Y} 
\end{align}
In Table \ref{vector-DS}, 
the Levi-Civita tensor $\epsilon^{mn}$ is defined as 
\begin{subequations}
\begin{align}
& \epsilon_{01} \ = \ - \epsilon^{01} \ = \ 1 \; , \ \ \ \epsilon^{mn} =
\ - \epsilon^{nm} \; , \\
& \gamma_m \gamma_n \ = \ \eta_{mn} + \epsilon_{mn} \gamma_3 \; , \ \
\ \epsilon_{mn} \epsilon_{kl} \ = \ - \eta_{mk} \eta_{nl} + \eta_{ml}
\eta_{nk} \; .
\end{align}
\end{subequations}
\begin{table}[htbp]
\begin{center}
\begin{tabular}{c||cccccc}
	$\Bracket{\cal F}{\cal G}$ & ${A_0}$ &
	${A_0^*}$ & ${\psi_0}$ &
	${\psi_0^c}$ & ${F_0}$ &
	${F_0^*}$ \\ \hline\hline
	${A_0^*}$ & $\frac{1}{8} (p^2 - 4 Y^2)$
	& $0$ & $0$ & $0$ & $0$ & $0$ \\ 
	${A_0}$ & $0$ & $\frac{1}{8} (p^2 - 4 Y^2)$
	& $0$ & $0$ & $0$ & $0$ \\ 
	${\ol{\psi}_0}$ & $0$ & $0$ & $\half
	(\Slash{p} + 2 Y)$ & $0$ & $0$ & $0$ \\ 
	${\ol{\psi_0^c}}$ & $0$ & $0$ & $0$ & $\half
	(\Slash{p} + 2Y)$ & $0$ & $0$ \\ 
	${F_0^*}$ & $0$ & $0$ & $0$ & $0$ & $\half $ &
	$0$ \\ 
	${F_0^*}$ & $0$ & $0$ & $0$ & $0$ & $0$ & $\half$
\end{tabular}
\caption{Two-point functions of component fields 
in the chiral superfield $\Phi_0$.}
\label{chiral-DS}
\end{center}
{\footnotesize
${\cal F}$ and ${\cal G}$ denote arbitrary fields. 
Multiplication by the coefficient $R(p^2)$ 
defined by Eq.~(\ref{R-Y}) is omitted in all components.
}
\end{table}
\begin{table}[htbp]
\begin{center}
\begin{tabular}{c||cccccc}
	$\Bracket{\cal F}{\cal G}$ & ${\lambda}$ &
	${\lambda^c}$ & ${D}$ &
	${M'}$ & ${N'}$ &
	${V_n}$ \\ \hline\hline
	${\ol{\lambda}}$ &$\half (\Slash{p} + 2 Y)$ &
	$0$ & $0$ & $0$ & $0$ & $0$ \\ 
	${\ol{\lambda^c}}$ & $0$ & $\half (\Slash{p} + 2
	Y)$ & $0$ & $0$ & $0$ & $0$ \\ 
	${D}$ & $0$ & $0$ & $\frac{1}{4}$ & $\half Y$
	& $0$ & $0$ \\
	${M'}$ & $0$ & $0$ & $\half Y$ & $\frac{1}{4}
	p^2$ & $0$ & $0$ \\ 
	${N'}$ & $0$ & $0$ & $0$ & $0$ & $\frac{1}{4}
	p^2$ & $\frac{i}{2} Y \epsilon_{nk} p^k$\\
	${V_m}$ & $0$ & $0$ & $0$ & $0$ &
	$-\frac{i}{2} Y \epsilon_{mk}p^k$ & $- \frac{1}{4} (\eta_{mn}
	p^2 - p_m p_n)$
\end{tabular}
\caption{Two-point functions of component fields 
in the vector superfield $V$.}
\label{vector-DS}
\end{center}
{\footnotesize
Note that
multiplication by the coefficient $R(p^2)$ 
defined by Eq.~(\ref{R-Y}) is omitted in all components.
}
\end{table}
In the diagramatics these two-point functions correspond to 
Feynman diagrams with external lines 
of auxiliary fields 
and loops of $N$ dynamical fields $A_i$ or $\psi_i$, 
which are listed in Figure 1.  
We find that the auxiliary fields 
become dynamical as bound states of 
the original dynamical fields. 

Since the fields $D$, $M'$, $N'$ and $V_m$ are not diagonal 
in two-point functions, as seen in Table \ref{vector-DS},   
we should diagonalize them in order to define propagators. 
If we redefine $D$ and $N'$ by 
\begin{subequations}
\begin{align}
 D' (p) \ &= \ D (p) - 2 Y M' (p) \; , \\
 N''(p) \ &= \ N' (p) 
       + 2i Y \frac{\epsilon_{mk} p^k}{p^2} V^m (p) \; , 
\end{align}
\end{subequations}
we can obtain the diagonal two-point 
functions with respect to the redefined fields 
$D'$ and $N''$, as listed in Table \ref{vector2-DS}.
\begin{table}[htbp]
\begin{center}
\begin{tabular}{c||cccccc}
	$\Bracket{\cal F}{\cal G}$ & ${\lambda}$ &
	${\lambda^c}$ & ${D'}$ &
	${M'}$ & ${N''}$ &
	${V_n}$ \\ \hline\hline
	${\ol{\lambda}}$ &$\half (\Slash{p} + 2 Y)$ &
	$0$ & $0$ & $0$ & $0$ & $0$ \\
	${\ol{\lambda^c}}$ & $0$ & $\half (\Slash{p} + 2
	Y)$ & $0$ & $0$ & $0$ & $0$ \\ 
	${D'}$ & $0$ & $0$ & $\frac{1}{4}$ & $0$ & $0$
	& $0$ \\
	${M'}$ & $0$ & $0$ & $0$ & $\frac{1}{4} ( p^2
	- 4 Y^2)$ & $0$ & $0$ \\ 
	${N''}$ & $0$ & $0$ & $0$ & $0$ & $\frac{1}{4}
	p^2$ & $0$ \\ 
	${V_m}$ & $0$ & $0$ & $0$ & $0$ & $0$ & $-
	\frac{1}{4} (p^2 - 4Y^2) \{\eta_{mn} - \frac{p_m p_n}{p^2}\} $
\end{tabular}
\caption{Diagonal two-point functions of component fields in 
the vector superfield $V$.}
\label{vector2-DS}
\end{center}
{\footnotesize
Note again that
multiplication by the coefficient $R(p^2)$ 
defined by Eq.~(\ref{R-Y}) is omitted in all components.
}
\end{table}


~From Tables \ref{chiral-DS} and \ref{vector2-DS}, 
we immediately find  
that all the fields, except for $F_0$, $D'$ and $N''$, 
acquire masses $m = |2Y|$.  
$N''$ is a massless Nambu-Goldstone field for
{\it chiral} $U(1)$ symmetry breaking, 
and $F_0$ and $D'$ remain auxiliary fields.  
We interpret these phenomena as follows.   
In order to keep the mass relation for ${\cal N}=2$ supersymmetry, 
the auxiliary field $D$ is mixed with $M'$, 
and $M'$ becomes massive with mass $|2Y|$.  
Also, the massless pseudo-scalar boson $N'$
is mixed with a massless gauge boson $V_m$.  
As a result, 
the gauge boson acquires a mass of $|2Y|$.
This phenomenon is known as the Schwinger mechanism~\cite{schwinger}. 
Therefore there appear no massless bosons,  
in agreement with Coleman's theorem.  
The auxiliary fields, $F_0$ and $D$, 
do not have physical massless poles. 
Since the Nambu-Goldstone boson $N'$ 
is absorbed into the gauge boson $V_m$, 
$N'$ also has no physical massless pole.

Let us discuss the high-energy behavior.
In the high-energy limit $p^2 \to \infty$, 
all the two-point functions of the
auxiliary fields are suppressed, 
because $R(p^2 \to \infty) \to 0$. 
This is consistent with the behavior of the auxiliary fields: 
The auxiliary fields propagate 
in the low-energy region as bound states, 
but they do not propagate and disappear in the high-energy region. 

Now, let us normalize all the fields properly. 
In the low-energy limit $p^2 \to 0$,  
we would like to obtain the normalized two-point functions. 
For example, we normalize $A_0 (x)$ by 
\begin{align}
& S_{\rm eff} \ = \ \int \! \frac{d^2 p}{(2 \pi)^2} \wt{A} (-p)
    \Pi_{A_0} (p) \wt{A} (p) + \cdots \ 
  = \ \int \! \frac{d^2 p}{(2 \pi)^2} \wt{A}'(-p) 
    \big\{ Z_A \Pi_{A_0} (p) \big\} \wt{A}' (p) + \cdots \; , \\
& Z_A \Pi_{A_0} (p) \ \ 
  \xrightarrow[p^2 \to 0] \ \ p^2 - 4 Y^2 \;, \\
& \frac{1}{8} Z_A R(p^2) \ \ \xrightarrow[p^2 \to 0] \ \ 1 \; , 
  \ \ \ Z_A \ = \ \frac{16 \pi Y^2}{N} \; , 
\end{align}
which fixes the renormalization constant $Z_A$.
We calculate the propagators for normalized fields 
in the next subsection.

\subsection{ Propagators and the $\mbf{\beta}$ function }
We now calculate propagators from Tables \ref{chiral-DS} and
\ref{vector2-DS}. 
In order to define the gauge field propagators, 
we introduce a covariant gauge fixing term with
a gauge parameter $\alpha$ as 
\begin{align}
\Bracket{V_m}{V_n} \ &\equiv \ - (p^2 - 4 Y^2) \left\{
\eta_{mn} - ( 1 - \alpha^{-1} ) \frac{p_m p_n}{p^2} \right\} \; .
\end{align}
We obtain the normalized propagators as   
\begin{subequations}
\begin{align}
& D_{A_0} (p) \ = \ \frac{1}{p^2 - 4 Y^2} \; , \ \ \ 
S_{\psi_0} (p) \ = \ \frac{1}{\Slash{p} + 2 Y} \; , \ \ \ 
D_{F_0} (p) \ = \ 1 \; , \\
& S_{\lambda} (p) \ = \ \frac{1}{\Slash{p} + 2 Y} \; , \ \ \
D_{D'} (p) \ = \ 1 \; , \ \ \
D_{M'} (p) \ = \ \frac{1}{p^2 - 4 Y^2} \; , \ \ \
D_{N''} (p) \ = \ \frac{1}{p^2} \; , \\
& D_V^{mn} (p) \ = \ - \frac{1}{p^2 - 4 Y^2} 
  \left\{\eta^{mn} - ( 1 - \alpha ) 
  \frac{p^m p^n}{p^2} \right\} \; ,
\end{align}
\end{subequations}
where we have expressed fields in question by 
the indices on $D$ and $S$. 

Before closing this section, we define the $\beta$ function. 
In order to do so, we introduce
a cutoff $\Lambda$ and a renormalization point $\mu$ as  
\begin{subequations}
\begin{align}
\frac{1}{g^2} \ &= \ \int \! \frac{d^2 k}{(2 \pi)^2 i} \frac{1}{-k^2 +
Y^2} \ = \ \frac{1}{4 \pi} \log \frac{\Lambda^2}{Y^2} \; , \\
\frac{1}{g_{\rm R}^2} \ &= \ \frac{1}{g^2} - \frac{1}{4 \pi} \log
\frac{\Lambda^2}{\mu^2} \ = \ \frac{1}{4 \pi} \log \frac{\mu^2}{Y^2}
\; .
\end{align}
\end{subequations}
Then we can define the $\beta$ function $\beta(g_{\rm R})$ by  
\begin{align}
\beta (g_{\rm R}) \ &= \ \lpd{\log \mu} \ g_{\rm R} \ = \ -
\frac{g_{\rm R}^3}{4 \pi} \ < \ 0 \; .
\end{align}
~From this equation,  
it can be found that the system of the Schwinger phase is 
asymptotically free.

\psfrag{Ai}{$A_i$}
\psfrag{A0}{$A_0$}
\psfrag{A0*}{$A_0^*$}
\psfrag{psi}{$\psi_i$}
\psfrag{psi0}{$\psi_0$}
\psfrag{lambda}{$\lambda$}
\psfrag{M}{$M$}
\psfrag{N}{$N$}
\psfrag{D}{$D$}
\psfrag{V}{$V_m$}

\vs{5}

\includegraphics[width=3.9cm]{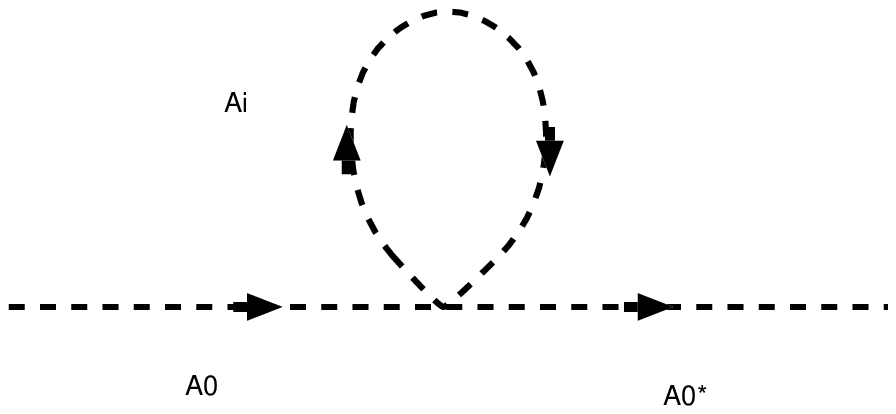} \ \ \ \  
\includegraphics[width=3.9cm]{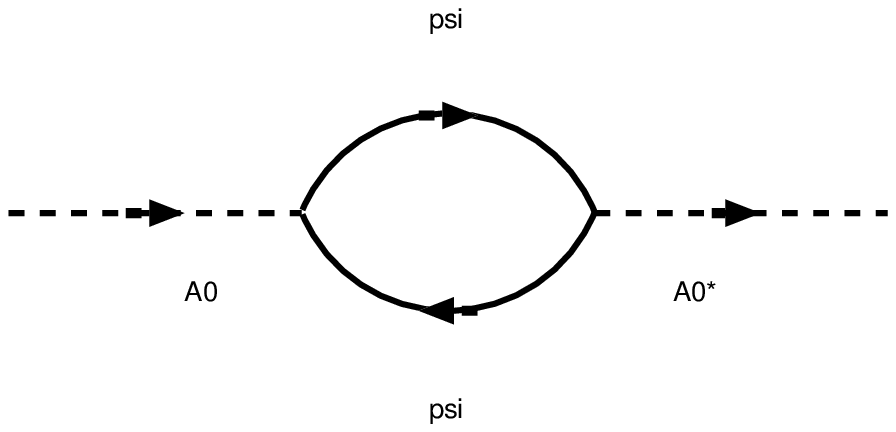}

\vs{5}

\includegraphics[width=3.9cm]{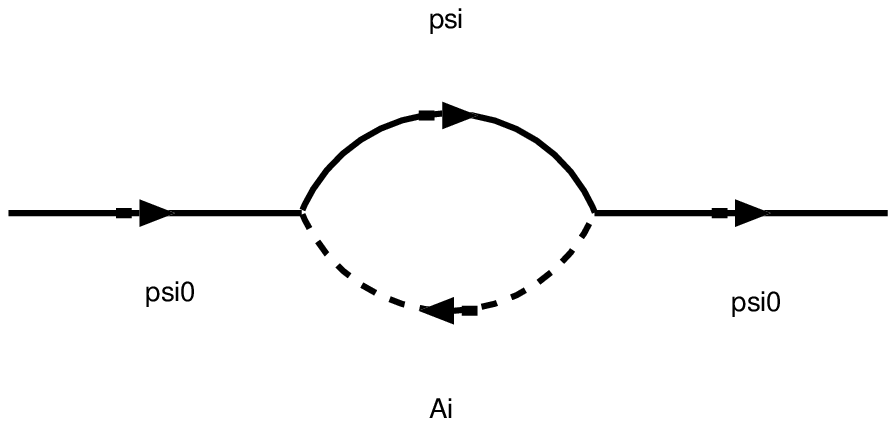} \ \ \ \  
\includegraphics[width=3.9cm]{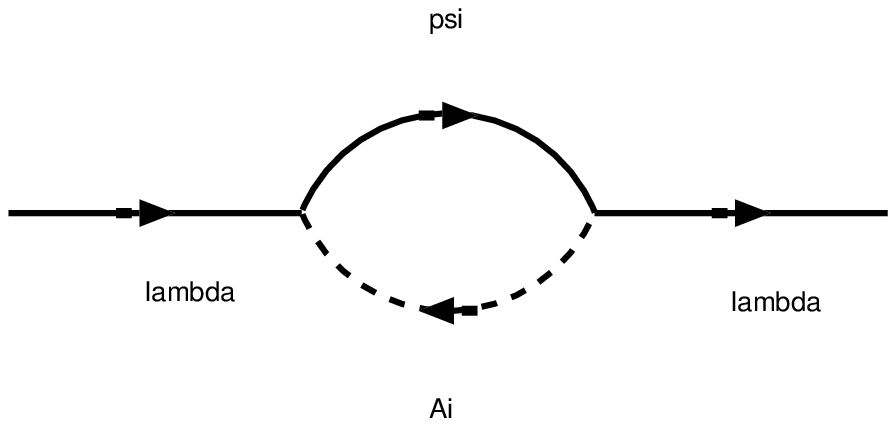}

\vs{5}

\includegraphics[width=3.9cm]{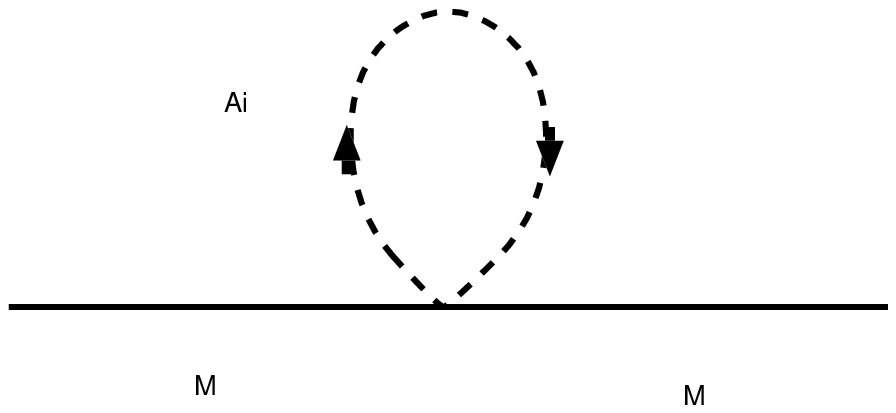}  \  
\includegraphics[width=3.9cm]{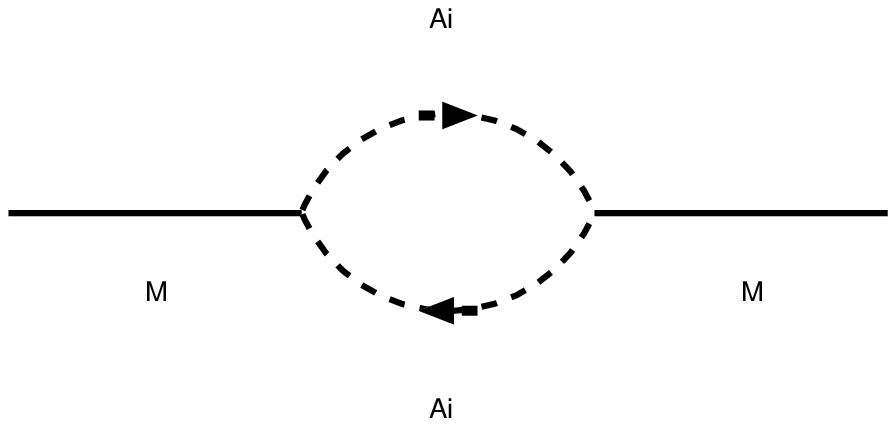}  \ 
\includegraphics[width=3.9cm]{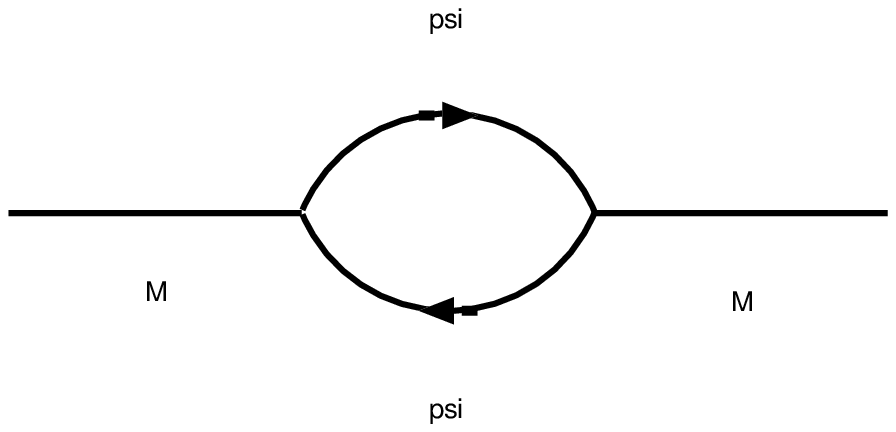}  \  
\includegraphics[width=3.9cm]{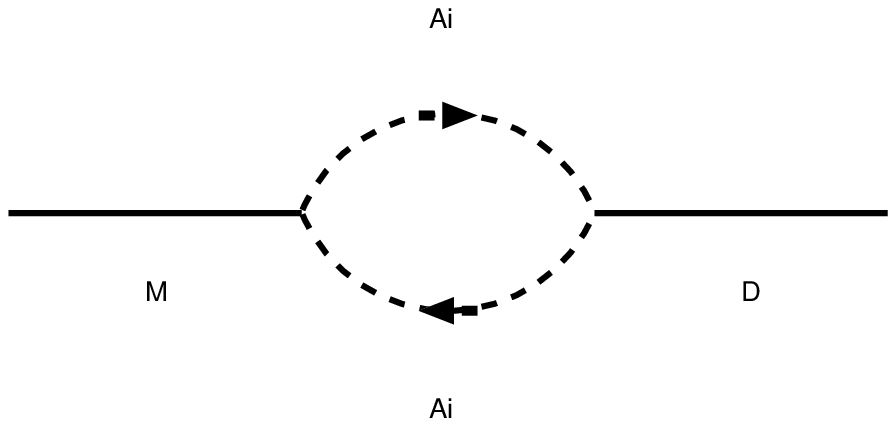}  \ 

\vs{5}

\includegraphics[width=3.9cm]{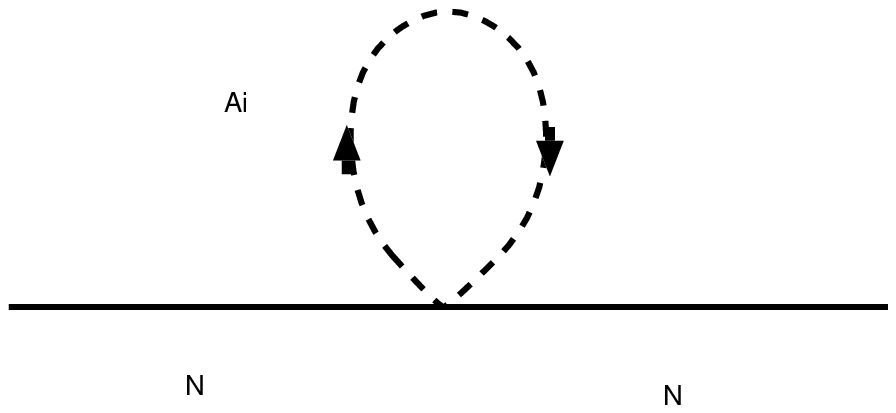} \ \  
\includegraphics[width=3.9cm]{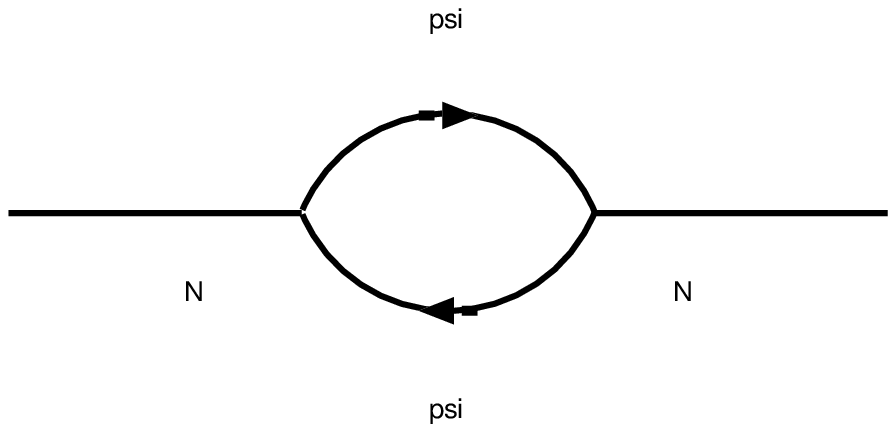} \ \ 

\vs{5}

\includegraphics[width=3.9cm]{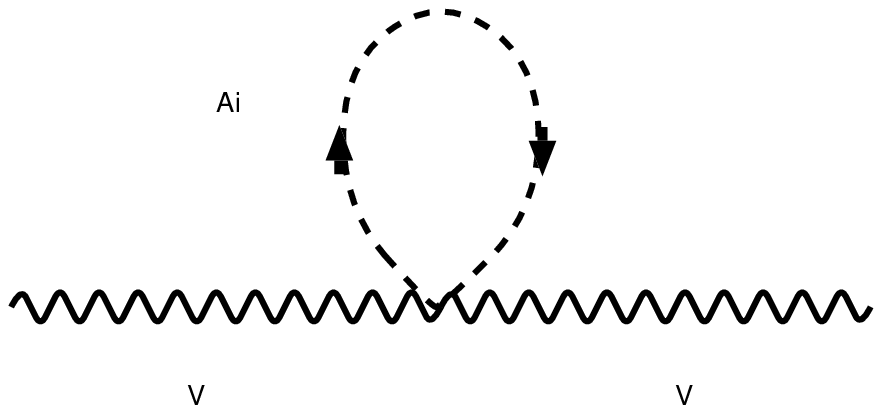}  \  
\includegraphics[width=3.9cm]{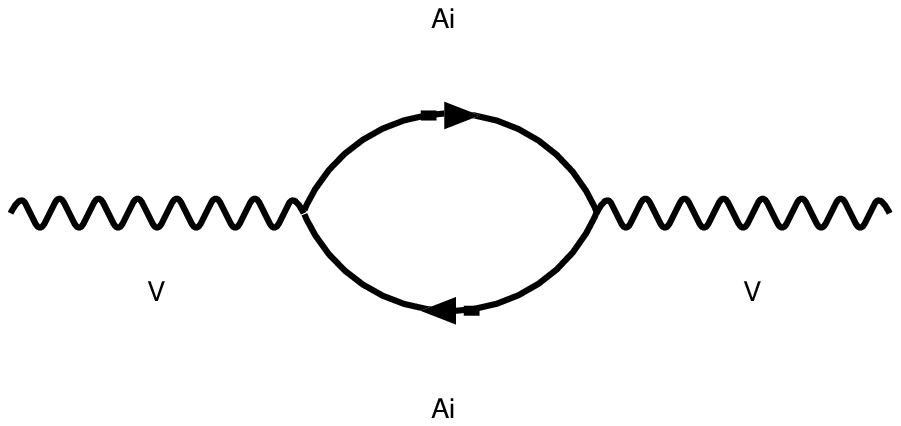}  \ 
\includegraphics[width=3.9cm]{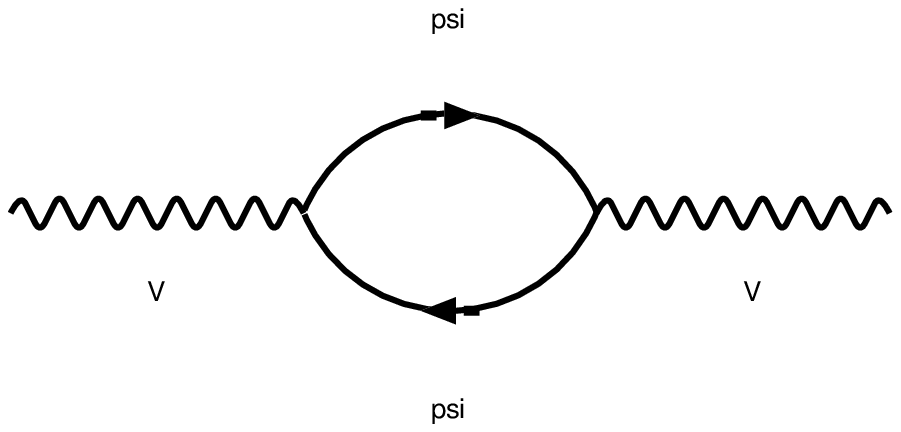}  \  
\includegraphics[width=3.9cm]{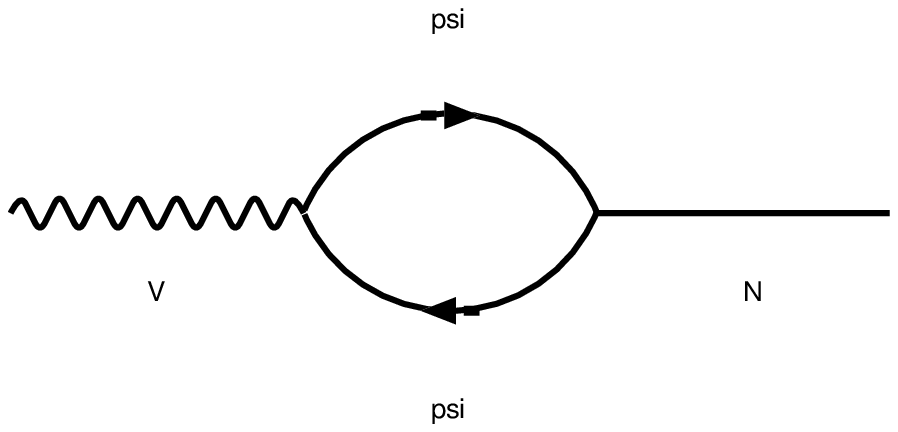}  \ 

\begin{center}
Figure 1: Feynman diagrams for two-point 
functions of auxiliary fields in the Schwinger phase.  
\end{center}
{\footnotesize 
Feynman diagrams with external lines,  
denoting auxiliary fields, 
and loops, denoting (integrated) dynamical fields 
$A_i$ and $\psi_i$, are listed.
All diagrams are order $N$.
We can read how auxiliary fields become 
bound states of original dynamical fields.
}

\section{ Higgs Phase } \label{Higgs-phase}
We discuss the Higgs phase in this section. 
This phase has to this time not been known to exist  
in two-dimensional nonlinear sigma models. 
It is quite different from the Schwinger phase.

\subsection{ Two-point functions }
The vacuum expectation value $X^2 =\phi_0^*\phi_0^{}$  
is non-zero, where $\left<A_0^{}(x)\right> = \phi_0^{}$.  
By using a phase rotation, 
$\phi_0^{}$ can be taken to be real: $\phi_0^{}=X$. 
We thus decompose the auxiliary fields 
$A_0^{}(x)$ and $A_0^*(x)$ into sums of 
$X$ and the quantum fluctuations by  
\begin{align}
  A_0(x) \ &= \ X + A_R(x) + i A_I(x) \; , \ \ \ 
  A_0^*(x) \ = \ X + A_R(x) - i A_I(x) \;. \label{VEV-Higgs}
\end{align} 
The other fields do not acquire vacuum expectation values. 
We again define $M'$ and $N'$ as quantum fluctuations of $M$ and $N$, 
and for the rest of the fields 
we use the same letters for the quantum fluctuations 
as those used for the original fields.    
By the vacuum expectation value (\ref{VEV-Higgs}), 
the global $U(1)$ symmetry and
the local $U(1)$ symmetry are broken down to
their linear combination 
in the last line of Table~\ref{symmetry-charge}.  
Since the unbroken symmetry is a global symmetry, 
the local $U(1)$ symmetry is broken.  
$A_I (x)$ becomes a (would-be) Nambu-Goldstone boson for 
the symmetry breaking of the local $U(1)$ symmetry. 
We can rewrite the Lagrangian as 
\begin{align}
{\cal L} \ &= \ - A_i^* \big[ \del^2 + X^2 \big] A_i + \half \Big( \
	\ol{\psi}_i \; , \;
	\ol{\psi_i^c} \ \Big) \left(
	\begin{array}{cc}
	i \gamma^m \del_m & - X \\
	- X & i \gamma^m \del_m
	\end{array} \right) \left(
	\begin{array}{c}
	\psi_i \\
	\psi_i^c
	\end{array} \right) \nonumber \\
& \ \ \ \ + V_m \Big( i A_i^* \del^m A_i^{} - \del^m A_i^* \cdot
	A_i^{} + \ol{\psi}_i \gamma^m \psi_i \Big) + V_m V^m
	A_i^* A_i^{} \nonumber \\
& \ \ \ \ + A_i^{} \Big( \ol{\lambda}
	\psi_i^c + \ol{\psi}_i \lambda^c \Big) +
	A_i^* \Big( \ol{\lambda^c} \psi_i^{} + \ol{\psi_i^c} \lambda
	\Big) \nonumber \\
& \ \ \ \ - M'{}^2 A_i^* A_i^{} + M' \Big( \ol{\psi}_i \psi_i^{} \Big)
	- N'{}^2 A_i^* A_i^{} - N' \Big( \ol{\psi}_i i \gamma_3
	\psi_i^{} \Big) - A_i^{} \Big( \ol{\psi_0^c} \psi_i^{} +
	\ol{\psi_i^c} \psi_0^{} \Big) - A_i^* \Big( \ol{\psi}_0
	\psi_i^c + \ol{\psi}_i \psi_0^c \Big) \nonumber \\
& \ \ \ \ - \Big( 2 X A_R^{} + A_R^2 + A_I^2 \Big) A_i^* A_i^{} -
	\Big( A_R^{} + i A_I^{} \Big) \ol{\psi_i^c} \psi_i^{} - \Big(
	A_R^{} - i A_I^{} \Big) \ol{\psi}_i \psi_i^c \nonumber \\
& \ \ \ \ + D A_i^* A_i^{} - \frac{N}{g^2} D + F_0^* A_i^2 + F_0^*
	A_i^*{}^2 \; . \label{Q-Higgs-L}
\end{align}
~From this equation, 
we find that the dynamical spinors $\psi_i$ acquire 
the Majorana mass terms 
$X \ol{\psi_i^c} \psi_i + X \ol{\psi}_i \psi_i^c$.
The auxiliary spinor $\psi_0$ also acquires  
the Majorana mass term, 
as seen in Table \ref{2-point-functions}, below.

Let us calculate the two-point functions in the Higgs phase. 
The definitions of the two-point functions are 
the same as those in the Schwinger phase. 
The coefficients of the two-point functions, $R(p^2)$, 
and the two-point functions of the gauge fields, $Q_{mn}(p)$,  
are defined by   
\begin{subequations}
\begin{align}
 & R(p^2) \ = \ \frac{N}{2 \pi} \int_0^1 \! dx \ 
     \frac{1}{X^2 - x(1-x) p^2}\; , \label{R} \\
 & Q_{mn}(p) \ = \ \left\{ X^2 \eta_{mn} - \frac{1}{4} 
     ( \eta_{mn}p^2 - p_m p_n ) \right\} R(p^2) 
    \ = \ \tilde Q_{mn}(p) R(p^2) \; .
\end{align}
\end{subequations}
We list all the two-point functions 
in Table~\ref{2-point-functions}, 
where we have again omitted 
the coefficient $R(p^2)$ for simplicity. 
\begin{sidetable}
  \begin{center}
    \begin{tabular}{c||cccccccccccc}
	$\langle {\cal F} {\cal G} \rangle$ & ${A_R}$
    & ${A_I}$ &
    ${\psi_0}$ & ${\psi_0^c}$ &
    ${F_0^*}$ & ${F_0}$ &
    ${D}$ & ${\lambda}$ &
    ${\lambda^c}$ & ${M'}$ &
    ${N'}$ & ${V_n}$ \\ \hline\hline
	${A_R}$ & $\frac{1}{4}p^2$ & $0$ & $0$ & $0$ &
    $0$ & $0$ & $-\half X$ & $0$ & $0$ & $0$ & $0$ & $0$ \\
 	${A_I}$ & $0$ & $\frac{1}{4}p^2$ & $0$ & $0$ &
    $0$ & $0$ & $0$ & $0$ & $0$ & $0$ & $0$ & $- \frac{i}{2} p_n X$ \\ 
	${\ol{\psi}_0}$ & $0$ & $0$ & $\half
    \Slash{p}$ & $0$ & $0$ & $0$ & $0$ & $0$ & $-X$ & $0$ & $0$ & $0$
    \\
 	${\ol{\psi_0^c}}$ & $0$ & $0$ & $0$ & $ \half
    \Slash{p}$ & $0$ & $0$ & $0$ & $-X$ & $0$ & $0$ & $0$ & $0$ \\
	${F_0}$ & $0$ & $0$ & $0$ & $0$ & $\half$ &
    $0$ & $0$ & $0$ & $0$ & $0$ & $0$ & $0$ \\  
 	${F_0^*}$ & $0$ & $0$ & $0$ & $0$ & $0$ &
    $\half$ & $0$ & $0$ & $0$ & $0$ & $0$ & $0$ \\ 
	${D}$ & $-\half X$ & $0$ & $0$ & $0$ & $0$ &
    $0$ & $\frac{1}{4}$ & $0$ & $0$ & $0$ & $0$ & $0$ \\
 	${\ol{\lambda}}$ & $0$ & $0$ & $0$ & $-X$ &
    $0$ & $0$ & $0$ & $ \half \Slash{p}$ & $0$ & $0$ & $0$ & $0$ \\
	${\ol{\lambda^c}}$ & $0$ & $0$ & $-X$ & $0$ &
    $0$ & $0$ & $0$ & $0$ & $ \half \Slash{p}$ & $0$ & $0$ & $0$ \\
 	${M'}$ & $0$ & $0$ & $0$ & $0$ & $0$ & $0$ &
    $0$ & $0$ & $0$ & $\frac{1}{4} (p^2 - 4 X^2)$ & $0$ & $0$ \\  
	${N'}$ & $0$ & $0$ & $0$ & $0$ & $0$ & $0$ &
    $0$ & $0$ & $0$ & $0$ & $\frac{1}{4} (p^2 - 4 X^2)$ & $0$ \\   
 	${V_m}$ & $0$ & $ \frac{i}{2} p_m X$ & $0$ &
    $0$ & $0$ & $0$ & $0$ & $0$ & $0$ & $0$ & $0$ & $\tilde Q_{mn}(p)$
    \end{tabular}
    \caption{The two-point functions in the Higgs phase.}
    \label{2-point-functions}
  \end{center}
{\footnotesize 
Multiplication by the coefficient $R(p^2)$ 
defined by Eq.~(\ref{R}) is omitted in all components.
}
\end{sidetable}
To diagonalize the two-point functions in Table~\ref{2-point-functions},
we should redefine fields by 
\begin{subequations}
\begin{align}
  A_I' (p) \ &= \ A_I (p) + \frac{2iX p^m}{p^2} V_m (p) \; , \ \ \ 
  D' (p) \ = \ D (p) - 2 X A_R (p) \; , \\
  \psi_0' (p) \ &= \ \psi_0 (p) + \lambda^c (p) \; , \ \ \ 
  \lambda' (p) \ = \ \lambda (p) - \psi_0^c (p) \; .
\end{align}
\end{subequations} 
The diagonal two-point functions 
are listed in Tables~\ref{Higgs-bosons} and \ref{psi-lambda}. 
(Note again that we have omitted 
the coefficient $R(p^2)$ in these tables.) 
Since other fields are already diagonal, 
they are not given in these tables.
The auxiliary fields, $F_0$ and $D$, 
do not have physical massless poles. 
The field $A_I$, which is the Nambu-Goldstone boson 
for local $U(1)$ symmetry breaking,  
is absorbed into the gauge boson $V_m$, and therefore 
it has no physical massless pole. 
\begin{table}[htbp]
\begin{center}
\begin{tabular}{c||cccc}
	$\Bracket{\cal F}{\cal G}$ & ${A_R}$ &
	${D'}$ & ${A_I'}$ &
	${V_n}$ \\ \hline\hline
	${A_R}$ & $\frac{1}{4} (p^2 - 4X^2)$ & $0$ &
	$0$ & $0$ \\
	${D'}$ & $0$ & $\frac{1}{4}$ & $0$ & $0$ \\
	${A_I'}$ & $0$ & $0$ & $\frac{1}{4} p^2$ & $0$ \\
	${V_m}$ & $0$ & $0$ & $0$ & $- \frac{1}{4}
	(p^2 - 4 X^2) \{\eta_{mn} - \frac{p_m p_n}{p^2} \}$
\end{tabular}
\caption{The two-point functions of $A_R$, $D'$, $A_I$, $V_m$.}
\label{Higgs-bosons}
\end{center}
\end{table}
\begin{table}[htbp]
\begin{center}
\begin{tabular}{c||cccc}
	$\Bracket{\cal F}{\cal G}$ & ${\psi_0'}$ &
	${\lambda^c{}'}$ & ${\cal
	F}{\psi_0^c{}'}$ & ${\lambda'}$ \\
	\hline\hline
	${\ol{\psi}_0'}$ & $\half (\Slash{p} + 2X)$ &
	$0$ & $0$ & $0$ \\
	${\ol{\lambda^c}'}$ & $0$ & $ \half (\Slash{p} - 2
	X)$ & $0$ & $0$ \\
	${\ol{\psi^c_0}'}$ & $0$ & $0$ & $ \half
	(\Slash{p} + 2X)$ & $0$ \\
	${\ol{\lambda}'}$ & $0$ & $0$ & $0$ & $- \half
	(\Slash{p} - 2X)$ \\
\end{tabular}
\caption{The two-point functions of $\psi_0'$ and $\lambda$.}
\label{psi-lambda}
\end{center}
\end{table}

We now discuss the high-energy behavior in the Higgs phase. 
In the high-energy limit $p^2 \to \infty$, 
all two-point functions of the auxiliary fields 
are suppressed, because $R (p^2 \to \infty) \to 0$,  
as in the Schwinger phase. 
We should normalize the auxiliary fields in the low-energy limit. 
For example, we normalize $A_R(x)$ by  
\begin{align}
 & S_{\rm eff} \ = \ \int \! \frac{d^2 p}{(2 \pi)^2} \wt{A}_R (-p)
      \Pi_{A_R} (p) \wt{A}_R (p) + \cdots \ 
  = \ \int \! \frac{d^2 p}{(2 \pi)^2} \wt{A}'_R (-p) 
    \big\{ Z_A \Pi_{A_R} (p) \big\} \wt{A}'_R (p) +\cdots \; , \\
& Z_A \Pi_{A_R} (p) \ \ \xrightarrow[p^2 \to 0] \ \ p^2 - 4 X^2 \; ,\\
  \ \ \ \ & \frac{1}{4} Z_A R(p^2) \ \ 
  \xrightarrow[p^2 \to 0] \ \ 1 \; , \ \ \ 
  Z_A \ = \ \frac{8 \pi X^2}{N} \; . 
\end{align}

Let us discuss the phenomena exhibited in this phase. 
In order to keep ${\cal N}=2$ supersymmetry, 
a massless boson $A_R$ is mixed with $D$, 
and then $A_R$ obtains mass $m = |2X|$. 
The Nambu-Goldstone boson $A_I$ is absorbed into 
a massless gauge boson $V_m$ to form 
a massive gauge boson with mass $m = |2X|$, 
as a result of the Higgs mechanism.   
Dirac fermions $\psi_0$ and $\lambda$ are mixed with each other 
and acquire masses $m = |2X|$.  
Since $F_0$ and $D'$ are independent of $p^2$ 
[but depend on $R(p^2)$], 
they remain auxiliary fields. 

To summarize, 
the original dynamical field have acquired mass $m=|X|$,  
and the $SO(N)$ symmetry is restored.  
In addition, all auxiliary fields, except for $F_0$ and $D'$, 
have become dynamical as bound states of 
the original dynamical fields, 
whose masses are all $m = |2X|$.

\subsection{ Propagators and the $\mbf{\beta}$ function }
In this subsection we calculate normalized propagators. 
We first introduce a covariant gauge fixing term with 
a gauge parameter
$\alpha$ in order to construct propagators of gauge fields by 
\begin{align}
\Bracket{V_m}{V_n} \ &\equiv \ - (p^2 - 4 X^2) \Big\{
\eta_{mn} - ( 1 - \alpha^{-1}) \frac{p_m p_n}{p^2} \Big\} \; .
\end{align}
We can calculate all of the normalized propagators. 
We have the following:  
\begin{subequations}
\begin{align}
 D_{M'} (p) \ &= \ \frac{1}{p^2 - 4 X^2} \; , \hspace{10mm}  
 D_{N'} (p) \ = \ \frac{1}{p^2 - 4 X^2} \; , \\
 S_{\psi_0'} (p) \ &= \ \frac{1}{\Slash{p} - 2 X} \; , 
                              \hspace{14mm}  
 S_{\lambda'} (p) \ = \ \frac{1}{\Slash{p} - 2 X} \; , \\
 D_{A_R} (p) \ &= \ \frac{1}{p^2 - 4 X^2} \; , \ \ \  D_{A_I'} (p) \
 = \ \frac{1}{p^2} \; , \\
 D_V^{mn} (p) \ &= \ - \frac{1}{p^2 - 4X^2} 
   \Big\{\eta^{mn} - ( 1 - \alpha ) 
   \frac{p^m p^n}{p^2} \Big\} \; , \\
 D_{D'} (p) \ &= \ 1 \; , \ \ \
 D_{F_0} (p) \ = \ 1 \; ,
\end{align}
\end{subequations}
where we have expressed fields in question 
by the indices on $D$ and $S$.

To define the $\beta$ function,  
we introduce a cutoff $\Lambda$ 
and a renormalization point $\mu$ as
\begin{subequations}
\begin{align}
\frac{1}{g^2} \ &= \ \int \! \frac{d^2 k}{(2 \pi)^2 i} 
   \frac{1}{-k^2 + X^2} \ 
 = \ \frac{1}{4 \pi} \log \frac{\Lambda^2}{X^2} \; , \\
  \frac{1}{g_{\rm R}^2} \ 
&= \ \frac{1}{g^2} - \frac{1}{4 \pi} \log \frac{\Lambda^2}{\mu^2} \ 
 = \ \frac{1}{4 \pi} \log \frac{\mu^2}{X^2}\; .
\end{align}
\end{subequations}
We thus define the $\beta$ function $\beta(g_{\rm R})$ by 
\begin{align}
  \beta (g_{\rm R}) \ &= \ \lpd{\log \mu} \ g_{\rm R} \ 
  = \ - \frac{g_{\rm R}^3}{4 \pi} \ < \ 0 \; , 
\end{align}
which shows that  
the system of the Higgs phase is also asymptotically free.


\psfrag{AR}{$A_R$}
\psfrag{AI}{$A_I$}

\vs{5}

\includegraphics[width=3.9cm]{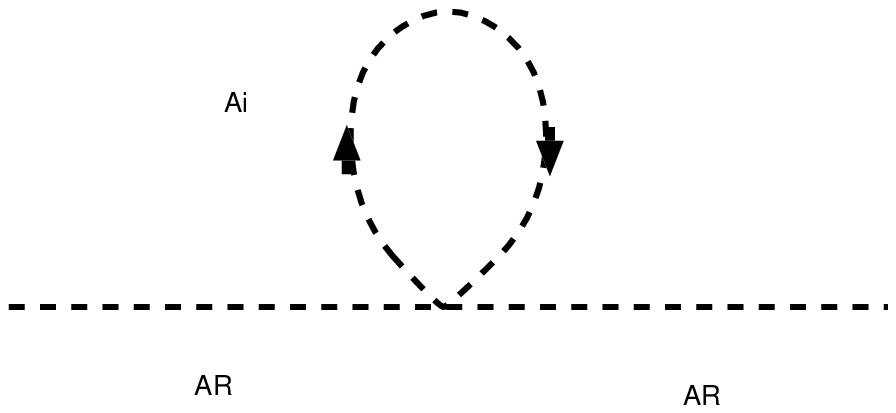}  \  
\includegraphics[width=3.9cm]{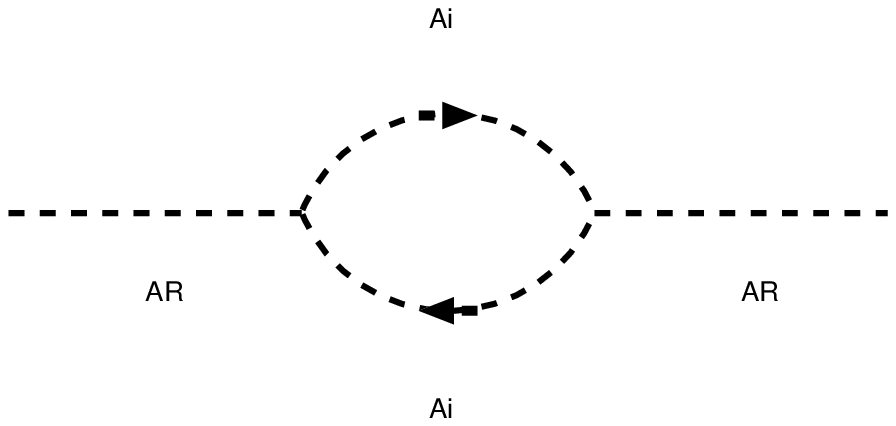}  \ 
\includegraphics[width=3.9cm]{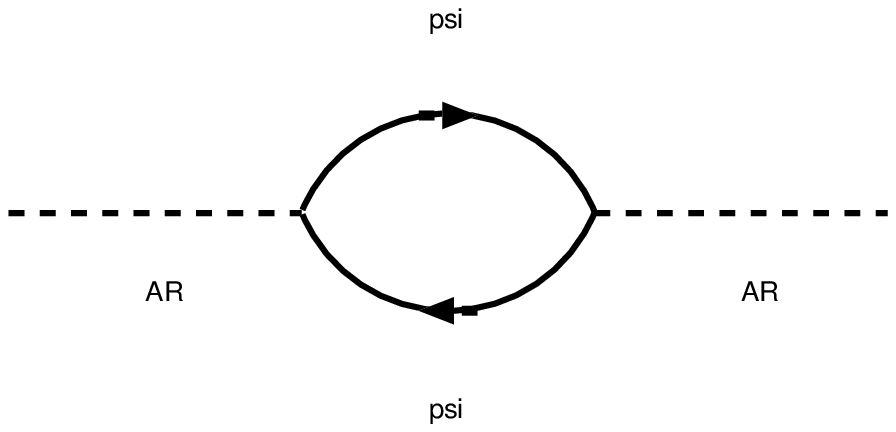}  \  
\includegraphics[width=3.9cm]{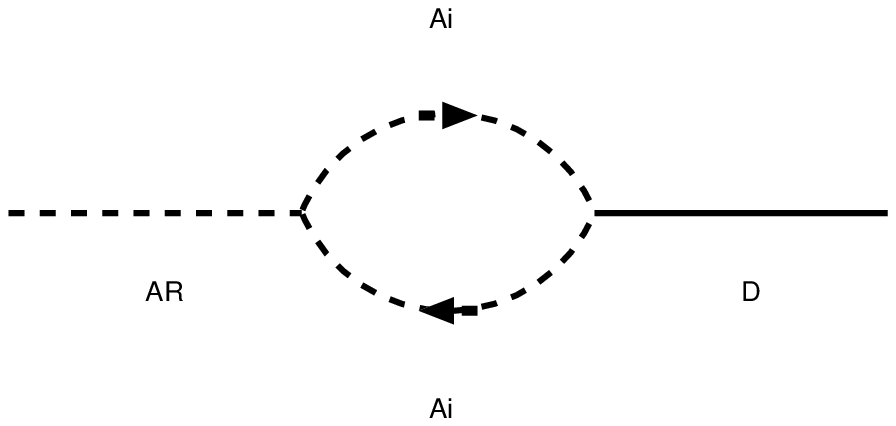}  \ 

\vs{5}

\includegraphics[width=3.9cm]{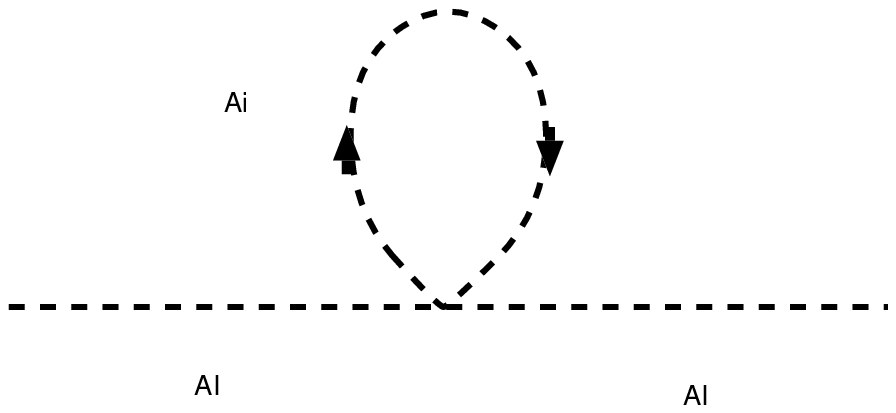}  \  
\includegraphics[width=3.9cm]{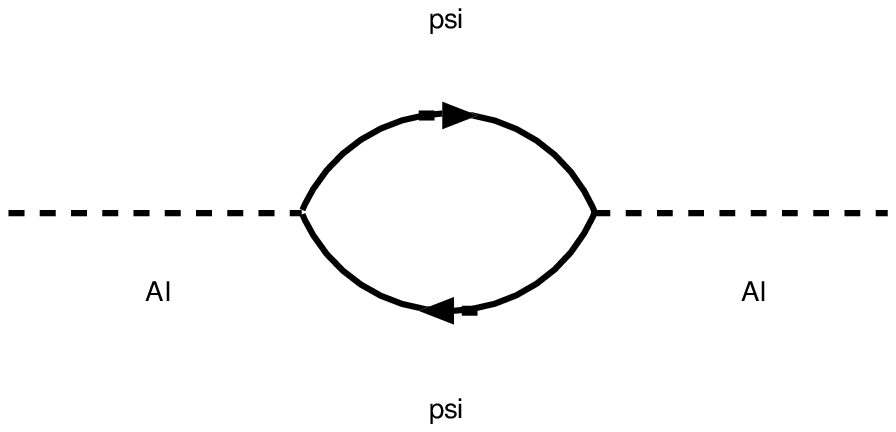}  \ 
\includegraphics[width=3.9cm]{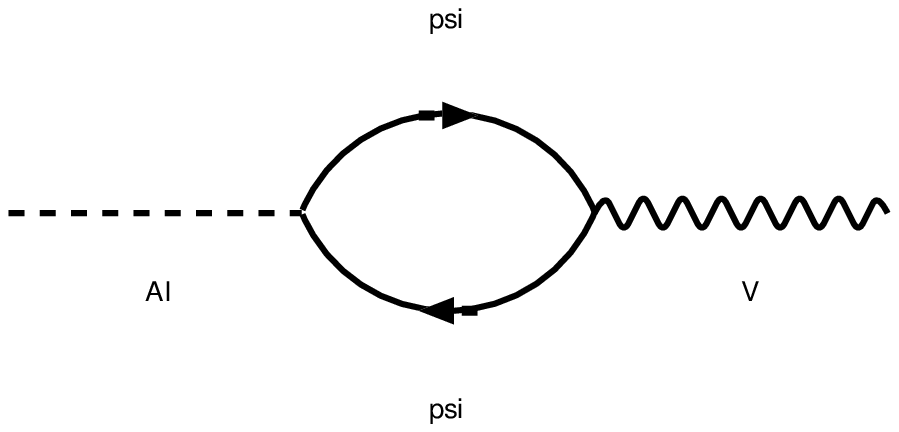}  \ 

\vs{5}

\includegraphics[width=3.9cm]{S-psi-1.eps} \ \  
\includegraphics[width=3.9cm]{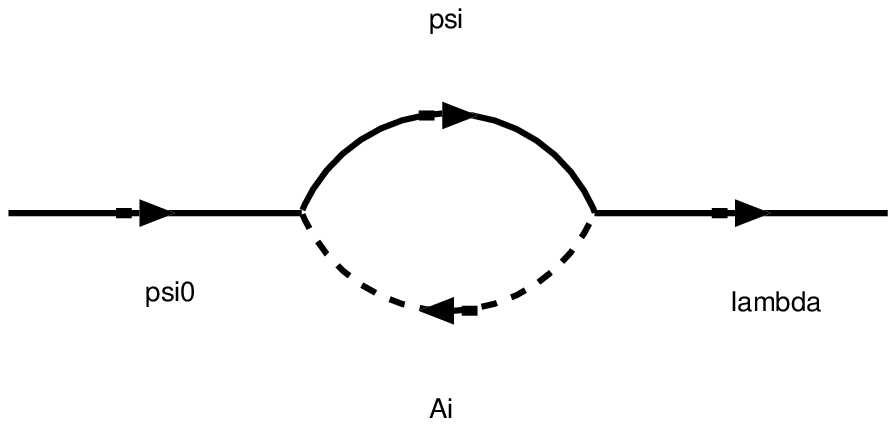} \ \  
\includegraphics[width=3.9cm]{S-lambda-1.eps} \ \ 

\vs{5}

\includegraphics[width=3.9cm]{S-M-1.eps} \ \  
\includegraphics[width=3.9cm]{S-M-3.eps} \ \ 

\vs{5}

\includegraphics[width=3.9cm]{S-N-1.eps} \ \  
\includegraphics[width=3.9cm]{S-N-2.eps} \ \ 

\vs{5}

\includegraphics[width=3.9cm]{S-V-1.eps}  \  
\includegraphics[width=3.9cm]{S-V-2.eps}  \ 
\includegraphics[width=3.9cm]{S-V-3.eps}  

\begin{center}
Figure 2: Feynman diagrams for two-point 
functions of auxiliary fields in the Higgs phase.  
\end{center}
{\footnotesize 
Feynman diagrams with external lines, 
denoting auxiliary fields, and loops, denoting (integrated) 
dynamical fields $A_i$ and $\psi_i$, are listed.
All diagrams are order $N$.
We can read how auxiliary fields become 
bound states of original dynamical fields.
}

\section{Conclusion and Discussion} \label{conclusion}
We have studied non-perturbative effects of 
the two-dimensional ${\cal N}=2$ supersymmetric 
nonlinear sigma model on the quadric surface 
$Q^{N-2}({\bf C})=SO(N)/SO(N-2)\times U(1)$ (the $Q^N$ model),  
by using auxiliary field and large-$N$ methods. 
To formulate the $Q^N$ model by auxiliary field methods, 
we needed two kinds of auxiliary superfields, 
a {\it vector} superfield $V(x,\theta,\thb)$ 
and a {\it chiral} superfield $\Phi_0(x,\theta,\thb)$. 
By integrating out the dynamical fields 
$A_i(x)$ and $\psi_i(x)$ [and $F_i(x)$], 
we calculated the effective potential. 
We have found that this model has two kinds of 
non-perturbatively stable vacua. 
In these vacua, scalar components of 
the auxiliary {\it vector} and {\it chiral} superfields,  
namely $M(x)$ [or $N(x)$] and $A_0(x)$, 
acquire non-zero vacuum expectation values, 
given by $Y$ and $X$, respectively.  
The former is the same vacuum as that of 
the ${\cal N}=2$ supersymmetric ${\bf C}P^{N-1}$ model. 
We call it the {\it Schwinger phase},  
since a massless gauge boson $V_m(x)$ 
becomes massive as a result of the Schwinger mechanism.
The latter is a new kind of vacuum, 
which has been seen here for the first time. 
We call it the {\it Higgs phase}, 
since a massless gauge boson becomes massive due to 
the Higgs mechanism. 

In the Schwinger phase, all component fields of 
the dynamical chiral superfields $\Phi_i(x,\theta,\thb)$, 
belonging to $SO(N)$ vectors, acquire masses $m=|Y|= \langle
M(x)\rangle$, 
and the $SO(N)$ symmetry is dynamically restored. 
In particular, Dirac spinors $\psi_i (x)$ acquire   
{\it Dirac} mass terms, 
which break the global {\it chiral} $U(1)$ symmetry spontaneously. 
Then, one of the auxiliary fields, $N (x)$, 
becomes a massless Nambu-Goldstone boson. 
In two dimensions, however, the appearance of 
a Nambu-Goldstone boson is forbidden by Coleman's theorem. 
The massless gauge field $V_m(x)$ and $N(x)$ are mixed,  
and the massless pseudo-scalar $N(x)$ is 
absorbed into the gauge boson as a result of the Schwinger mechanism.
In addition, the auxiliary field $M(x)$ 
becomes massive through mixing with $D(x)$ 
to preserve supersymmetry. 
Therefore, all massless bosons disappear from 
the physical spectrum 
in agreement with Coleman's theorem. 
Then, all the component fields of the auxiliary superfields, 
except for auxiliary fields needed for supersymmetry, 
acquire masses $m = |2Y|$, and supersymmetry is preserved.

In the Higgs phase, all component fields of 
the dynamical superfields, 
belonging to the $SO(N)$ vector representation, 
obtain masses $m = |X|$, 
and the $SO(N)$ is again dynamically restored. 
Here, the complex fermions $\psi_i(x)$ acquire  
{\it Majorana} mass terms. 
By these mass terms, 
the $U(1)_{\rm global} \times U(1)_{\rm local}$ symmetry 
is broken to their linear combination.   
In this phase, the imaginary part of the scalar field $A_0(x)$,  
$A_I(x)$, becomes a (would-be) Nambu-Goldstone boson.  
This massless boson, however, is absorbed into 
the gauge boson $V_m (x)$ to form a massive gauge boson, 
as a result of the Higgs mechanism. 
The real part of $A_0(x)$, $A_R(x)$, 
becomes massive through mixing with $D(x)$.
Therefore, all massless bosons again disappear from 
the physical spectrum 
in agreement with Coleman's theorem. 
All component fields of the auxiliary superfields, 
except for auxiliary fields for supersymemtry, 
acquire masses $m = |2X|$, and supersymmetry is preserved. 

Furthermore we have shown that both phases are asymptotically free 
by calculating the $\beta$ functions.

\vs{5}

In this paper, we have discussed 
the leading order of the $1/N$ expansion. 
It is interesting to consider the next order of this model,  
in particular whether there exist next-to-leading 
order corrections of the $1/N$ expansion.
Scalar components of auxiliary vector and chiral superfields  
acquire non-zero vacuum expectation values  
in the Schwinger and Higgs phases, respectively.  
They are similar to the Coulomb and Higgs branches of
four-dimensional ${\cal N}=2$ supersymmetric QCD 
in the sense that scalar components of gauge multiplets 
and hyper-multiplets acquire non-zero 
vacuum expectation values.
Further investigation of the similarities 
to four-dimensional ${\cal N}=2$ supersymmetric QCD 
would be interesting. 

Let us now discuss possible generalizations of this model.  
We would like to discuss non-perturbative effects of 
nonlinear sigma models on 
other Hermitian symmetric spaces summarized in 
Table \ref{table-HSS} in the Introduction. 
For example, non-Abelian gauge bosons would be dynamically 
generated in the Grassmannian model, 
the $SO(2N)/U(N)$ model, and the $Sp(N)/U(N)$ model.  
It is also an interesting task to generalize this model to 
three dimensions. 
In three dimensions, nonlinear sigma models are perturbatively 
non-renormalizable, 
but they are renormalizable in the $1/N$ expansion.
Recently there has been progress in 
the study of ${\cal N}=2$ (${\cal N}=4$) 
supersymmetric nonlinear sigma 
models on ${\bf C}P^{N-1}$ 
(the cotangent bundle over ${\bf C}P^{N-1}$)  
in three dimensions~\cite{3dim}. 
By dimensional reduction of our model in 
four dimensions~\cite{HN1} to three dimensions, 
we would be able to treat other models in three dimensions. 
We hope that the investigation of 
these new models in two (or three) dimensions 
would provide us further understanding of  
non-perturbative aspects of quantum field theories.

\section*{Acknowledgements} 

We would like to thank Nobuyoshi Ohta for useful comments. 
T.~K. would like to thank Naoto Yokoi for several
discussions about supersymmetric vacua. 
M.~N. is grateful to Takeo Inami and 
Masayoshi Yamamoto for valuable discussions and comments.  
The work of M.~N. is supported in part 
by JSPS Research Fellowships.

\begin{appendix}

\section{ Notation for $\mbf{{\cal N}=1}$ Supersymmetry in 
Four Dimensions}\label{four-dim}
Before constructing ${\cal N}=2$ supersymmetry in two dimensions in 
the next appendix, 
we here define the notation for the spinors and 
${\cal N}=1$ supersymmetry in 
four dimensions~\cite{WessBagger}. 
The space-time metric is
$\eta_{\mu \nu} \ = \ {\rm diag.} ( - + + + )$. 
The Majorana spinors $\psi^M$ can be written by 
the Weyl spinors as
\begin{align}
 \ \psi^M \ = \ \left( 
	\begin{array}{c}
		\psi_{\alpha} \\
		\ol{\psi}^{\dot{\alpha}}
	\end{array} \right) \,, \hspace{10mm} 
 \ol{\psi^M} \ = \ \Big( -
	\psi^{\alpha} ,
	- \ol{\psi}_{\dot{\alpha}} \Big) \;. 
\end{align}
The Dirac matrices in four dimensions are
\begin{align}
  \ \gamma^{\mu} \ = \ \left(
	\begin{array}{cc}
		0 & \dps (\sigma^{\mu})_{\alpha \dot{\beta}} \\
		\dps (\ol{\sigma}^{\mu})^{\dot{\alpha} \beta} & 0 
	\end{array} \right) \, , \ \ \sigma^0 \ = \ \left(
	\begin{array}{cc}
		-1 & 0 \\
		0 & -1 
	\end{array} \right) \, , \ \ 
\gamma_5 \ = \ \gamma^0 \gamma^1 \gamma^2 \gamma^3 \ = \ \left(
	\begin{array}{cc}
		-i & 0 \\
		0 & i
	\end{array} \right) \, , 
\end{align}
where $\sigma^i$ are the Pauli matrices:  
$\ol{\sigma}^0 \ = \ \sigma^0 \; , 
\ \ \ol{\sigma}^i \ = \ - \sigma^i$.
We note the identity
\begin{align} 
 & \ \psi \sigma^{\mu} \ol{\chi} \ 
 = \ - \ol{\chi} \, \ol{\sigma}^{\mu} \psi \; , \ \ \ \ol{\psi} \ 
 = \ \psi^{\dagger} 
\end{align}
for the Weyl spinors. 

We now give the notation for superfields.  
A chiral superfield, satisfying 
$\ol{D}_{\dot\alpha}\phi(x,\theta,\thb)=0$, is 
\begin{align}
\phi (y, \theta) \ &= \ A (y) + \sqrt{2} \theta \psi (y) + \theta
    \theta F (y), \nonumber \\
\phi (x,\theta,\thb)\ &= 
   \ A (x) + i \theta \sigma^{\mu} \ol{\theta} \del_{\mu} A (x) 
  + \frac{1}{4} \theta \theta \ol{\theta} \ol{\theta} \square A (x) 
  + \sqrt{2} \theta \psi (x) - \frac{i}{\sqrt{2}} \theta \theta 
    \del_{\mu} \psi (x) \sigma^{\mu} \ol{\theta} 
  + \theta \theta F (x) \; ,
\end{align}
where $y^{\mu} = x^{\mu} + i \theta \sigma^{\mu}\ol{\theta}$. 
A vector superfield, satisfying 
$V(x,\theta,\thb)^{\dagger}=V(x,\theta,\thb)$, is
\begin{align}
V (x , \theta, \ol{\theta}) \ &= \ - \theta \sigma^{\mu} \ol{\theta}
V_{\mu} + i \theta \theta \ol{\theta} \ol{\lambda} (x) - i \ol{\theta}
\ol{\theta} \theta \lambda (x) + \half \theta \theta \ol{\theta}
\ol{\theta} D (x) \; 
\end{align}
in the Wess-Zumino gauge.
To perform the dimensional reduction, 
we must consider the Fermion bilinear forms  
\begin{subequations}
\begin{align}
\ol{\psi_i^M} \psi_i^M \ &= \ - \psi_i \psi_i - \ol{\psi}_i
\ol{\psi}_i \; , \\
\ol{\psi_i^M} \gamma_5 \psi_i^M \ &= \ i \psi_i \psi_i - i \ol{\psi}_i
\ol{\psi}_i \; , \\
\ol{\psi_i^M} \gamma^{\mu} \psi_i^M \ &= \ - \psi_i \sigma^{\mu}
\ol{\psi}_i - \ol{\psi}_i \ol{\sigma}^{\mu} \psi_i \ = \ 0 \; , \\
\ol{\psi_i^M} \gamma^{\mu} \gamma_5 \psi_i^M \ &= \ - i \psi_i
\sigma^{\mu} \ol{\psi}_i + i \ol{\psi}_i \ol{\sigma}^{\mu} \psi_i \ =
\ 2 i \ol{\psi}_i \ol{\sigma}^{\mu} \psi_i \; , \\
\ol{\psi_i^M} \gamma^{\mu} \del_{\mu} \psi_i^M \ &= \ - \psi_i
\sigma^{\mu} \del_{\mu} \ol{\psi}_i - \ol{\psi}_i \ol{\sigma}^{\mu}
\del_{\mu} \psi_i \ = \ - 2 \ol{\psi}_i \ol{\sigma}^{\mu} \del_{\mu}
\psi_i \; , \\
\ol{\psi_i^M} \gamma^{\mu} \gamma_5 \del_{\mu} \psi_i^M \ &= \ i
\psi_i \sigma^{\mu} \del_{\mu} \ol{\psi}_i - i \ol{\psi}_i
\ol{\sigma}^{\mu} \del_{\mu} \psi_i \ = \ 0 \; .
\end{align}
\end{subequations}
Here integration over $x$ is implied for each equation.

\section{ Dimensional Reduction to Two Dimensions} 
\label{dimensional-reduction} 
Before dimensional reduction,  
we change some notation. 
First, we change 
the sign of the space-time metric according to
\begin{align}
  \eta_{\mu \nu} \ = \ {\rm diag.} ( - + + + ) \ 
 = \ - {\rm diag.} ( + - - - ) \ 
 = \ - \wt{\eta}_{\mu \nu}.
\end{align}
Next, we change the Dirac gamma matrices $\gamma^{\mu}$ 
from those in Appendix~A to ours:
\begin{subequations}
\begin{align}
& \gamma^{\mu} \ = \ \Gamma^{\mu} \; , \ \ i \gamma_5 \ = \ i \gamma^0
\gamma^1 \gamma^2 \gamma^3 \ = \ i \Gamma^0
\Gamma^1 \Gamma^2 \Gamma^3 \ = \ \Gamma_5 \; , \\
& \{ \gamma^{\mu} , \gamma^{\nu} \} \ = \ - 2 \eta^{\mu \nu} \ = \ 2
\wt{\eta}^{\mu \nu} \ = \ \{ \Gamma^{\mu} , \Gamma^{\nu} \} \; .
\end{align}
\end{subequations}

We denote the Dirac matrices in four dimensions and 
in two dimensions  
by $\Gamma^{\mu}$ and $\gamma^m$, respectively. 
The latter can be embedded in the former as follows:
\begin{subequations}
\begin{align}
& \Gamma^{m} \ = \ \gamma^{m} \otimes \sigma_1 \ = \ \left(
	\begin{array}{cc}
	0 & \gamma^{m} \\
	\gamma^{m} & 0 
	\end{array} \right) \; , \ \ m = 0, 1 \; , \\
& \Gamma^2 \ = \ i \gamma_3 \otimes \sigma_1 \ = \ \left(
	\begin{array}{cc}
	0 & i \gamma_3 \\
	i \gamma_3 & 0 
	\end{array} \right) \; , \ \ \Gamma^3 \ = \ {\bf 1} \otimes i 
	\sigma_2 \ = \ \left(
	\begin{array}{cc}
	0 & {\bf 1} \\
	- {\bf 1} & 0 
	\end{array} \right) \; , \\
& \Gamma_5 \ = \ {\bf 1} \otimes \sigma_3 \; , \ \ C_4 \ = \ i
	\Gamma^1 \Gamma^2 \ = \ \gamma^0 \otimes {\bf 1} \ = \ \left(
	\begin{array}{cc}
	\gamma^0 & 0 \\
	0 & \gamma^0
	\end{array} \right) \ = \ - C_2 \otimes {\bf 1} \; , \\
& \gamma^0 \ = \ \sigma_2 \; , \ \ \gamma^1 \ = \ i \sigma_1 \; , \ \
	\gamma_3 \ = \ \gamma^0 \gamma^1 \ = \ \sigma_3 \; , \\
& C_2 \ = \ - \gamma^0 \; , \\
& C_2 \ = \ - C_2^T \ = \ - C_2^* \ = \ C_2^{\dagger} \ = \ C_2^{-1}
	\; , \\
& C_2^{-1} \gamma^{\mu} C_2 \ = \ - \gamma^{\mu}{}^T \; , \ \ C_2^{-1}
	\gamma_3 C_2 \ = \ - \gamma_3^T \; .
\end{align}
\end{subequations}

The four-dimensional Majorana spinor $\psi^M$ 
can be expressed by the two-component Weyl spinor $\psi$ as  
$\psi^M \ = \ \left( \begin{array}{c}
	\psi \\
	\psi'
	\end{array} \right)$. 
Then the Majorana condition $\psi^M = C_4 \ol{\psi^M}^T$ becomes
\begin{align}
\left( \begin{array}{c}
	\psi \\
	\psi' 
	\end{array} \right) \ &= \ \left( \begin{array}{c}
	- C_2 \ol{\psi}'{}^T \\
	- C_2 \ol{\psi}^T
	\end{array} \right) \ 
 = \ \left( \begin{array}{c}
	\psi \\
	- \chi
	\end{array} \right) \,, \hspace{10mm}
  \chi \ = \ C_2 \ol{\psi}^T \ = \ \psi^* \; . 
\end{align}
If the Majorana spinor $\psi^M$ in four dimensions 
does not depend on $x^2$ and $x^3$, 
we can rewrite this spinor with 
the two-dimensional Dirac spinor $\psi$. 
The two-component Weyl spinors $\psi$ and $\chi$ in four dimensions 
become Dirac spinors in two dimensions 
when we apply dimensional reduction 
from $\{x^{\mu}\}$ to $\{x^m\}$ 
($\mu=0,1,2,3$; $m = 0,1$). 
Fermion bilinear forms become 
\begin{subequations}
\begin{align}
\ol{\psi_i^M} \psi_j^M \ &= \ - \ol{\psi_i^c} \psi_j - \ol{\psi}_i
\psi_j^c \; , \\
\ol{\psi_i^M} \Gamma_5 \psi_j^M \ &= \ - \ol{\psi_i^c} \psi_j +
\ol{\psi}_i \psi_j^c \; , \\
\ol{\psi_i^M} \Gamma^m \psi_i^M \ &= \ \ol{\psi_i^M} \Gamma^2 \psi_i^M
\ = \ \ol{\psi_i^M} \Gamma^3 \psi_i^M \ = \ 0 \; , \\
\ol{\psi_i^M} \Gamma^m \Gamma_5 \psi_i^M \ &= \ 2 \ol{\psi}_i \gamma^m
\psi_i \; , \\
\ol{\psi_i^M} \Gamma^2 \Gamma_5 \psi_i^M \ &= \ 2 i \ol{\psi}_i \gamma_3
\psi_i \; , \\
\ol{\psi_i^M} \Gamma^3 \Gamma_5 \psi_i^M \ &= \ -2 \ol{\psi}_i \psi_i
\; , \\
\ol{\psi_i^M} \Gamma^m \del_m \psi_i^M \ &= \ 2 \ol{\psi}_i \gamma^m
\del_m \psi_i \; , \\
\ol{\psi_i^M} \Gamma^2 \del_2 \psi_i^M \ &= \ 2i \ol{\psi}_i \gamma_3
\del_2 \psi_i \; , \\
\ol{\psi_i^M} \Gamma^3 \del_3 \psi_i^M \ &= \ - 2 \ol{\psi}_i \del_3
\psi_i \; , \\
\ol{\psi_i^M} \Gamma^{\mu} \Gamma_5 \del_{\mu} \psi_i^M \ &= \ 0 \; ,
\end{align}
\end{subequations}
where the integral over $x$ for each equation is implied. 

\end{appendix}

}




\begin{thebibliography}{99}
\bibitem{SUSY-QCD}
N.~Seiberg, 
Nucl. Phys. {\bf B435} (1995) 129; 
N.~Seiberg and E.~Witten,  
Nucl. Phys. {\bf B426} (1994) 19, hep-th/9407087, 
Erratum-ibid. {\bf B430} (1994) 485;
Nucl. Phys. {\bf B431} (1994) 484, hep-th/9408099;\\  
P.~C.~Argyres, M.~R.~Plesser and N.~Seiberg, 
Nucl. Phys. {\bf B471} (1996) 159, hep-th/9603042.  

\bibitem{Polyakov}
A. M. Polyakov, ``Gauge Fields and Strings'', 
Harwood Academic Publishers (1987).

\bibitem{Coleman}
S.~Coleman, ``Aspects of Symmetry'', 
Cambridge University Press (1985).

\bibitem{Abdalla}
E.~Abdalla, M.~C.~B.~Abdalla and K.~D.~Rothe, 
``Non-perturbative Methods in 2-Dimensional Quantum Field Theory'', 
World Scientific (1991).

\bibitem{Hi}
K.~Higashijima, Prog. Theor. Phys. Suppl. {\bf 104} (1991) 1. 

\bibitem{DLD}
A.~D'adda, M.~L\"{u}scher and P.~Di~Vecchia, 
Nucl. Phys. {\bf B146} (1978) 63.

\bibitem{WiAl} 
E.~Witten, 
Phys. Rev. {\bf D16} (1977) 2991; 
O.~Alvarez, Phys. Rev. {\bf D17} (1978) 1123.

\bibitem{Wi}
E.~Witten, Nucl. Phys. {\bf B149} (1979) 285.

\bibitem{DDL}
A.~D'adda, P.~Di~Vecchia and M.~L\"{u}scher, 
Nucl. Phys. {\bf B152} (1979) 125.

\bibitem{Ao}
S.~Aoyama, Nuovo Cim. {\bf 57A} (1980) 176.

\bibitem{Zu}
B.~Zumino,
Phys. Lett. {\bf 87B} (1979) 203.

\bibitem{HN1}
K.~Higashijima and M.~Nitta, 
Prog. Theor. Phys. {\bf 103} (2000) 635, hep-th/9911139.

\bibitem{HN2}
K.~Higashijima and M.~Nitta, 
Prog. Theor. Phys. {\bf 103} (2000) 833, hep-th/9911225. 

\bibitem{HN3}
K.~Higashijima and M.~Nitta, 
``Supersymmetric Nonlinear Sigma Models'', 
hep-th/0006025, to appear in Proceedings of 
Confinement 2000 held at Osaka, Japan, March, 2000;\\  
``Auxiliary Field Formulation of 
Supersymmetric Nonlinear Sigma Models'', 
hep-th/0008240, to appear in Proceedings of ICHEP 2000 
held at Osaka, Japan, July, 2000.  

\bibitem{IKK}
K.~Itoh, T.~Kugo and H.~Kunitomo,
Nucl. Phys. {\bf B263} (1986) 295.

\bibitem{Ni}
M.~Nitta, ``K\"{a}hler Potential for 
Global Symmetry Breaking in
Supersymmetric Theories'', hep-th/9903174.

\bibitem{Co}
S.~Coleman, Comm. Math. Phys. {\bf 31} (1973) 259.

\bibitem{KN} 
S.~Kobayashi and K.~Nomizu, 
``Foundations of Differential Geometry Volume II'', 
Wiley Interscience (1996).

\bibitem{DV}
F.~Delduc and G.~Valent, 
Nucl. Phys. {\bf B253} (1985) 494.

\bibitem{schwinger}
J.~Schwinger, Phys. Rev. {\bf 125} (1962) 397, {\bf 130} (1962) 2425.

\bibitem{3dim}
T.~Inami, Y.~Saito and M.~Yamamoto, 
Prog. Theor. Phys. {\bf 103} (2000) 1283, hep-th/0003013;\\   
Phys. Lett. {\bf B495} (2000) 245, hep-th/0008195. 
 
\bibitem{WessBagger}
J.~Wess and J.~Bagger, ``Supersymmetry and Supergravity (2nd ed.)'',
Princeton University Press (1992).

\end{thebibliography}
\end{document}